\documentclass[reprint,secnumarabic,amssymb, nobibnotes, aps, prd,superscriptaddress]{revtex4-2}

\usepackage[none]{hyphenat}
\usepackage{graphicx}
\usepackage{xcolor}
\usepackage{scalerel}
\usepackage{mathtools} 
\usepackage{amsmath}
\usepackage{physics}
\definecolor{bluecitation}{RGB}{45, 48, 146}
\usepackage[colorlinks=true,allcolors=bluecitation]{hyperref}%

\newcommand\fp{$3s^{-1}4p$ } 
\newcommand\td{$3s^{-1}3d$ } 
\newcommand\ff{$3s^{-1}4f$ } 

\newcommand\tdlis{\mbox{$3s^{-1}3d\,{\otimes}\,(N{-}1)\gamma_{\mathrm{IR}}$}} 
\newcommand\fslis{\mbox{$3s^{-1}5s\,{\otimes}\,(N{-}1)\gamma_{\mathrm{IR}}$}} 
\newcommand\fflis{\mbox{$3s^{-1}4f\,{\otimes}\,(N{-}2)\gamma_{\mathrm{IR}}$}} 

\begin{document}


\title{Multi-polariton control in attosecond transient absorption of autoionizing states }%


\author{S. Yanez-Pagans}%
\affiliation{Department of Physics, University of Arizona, Tucson, Arizona 85721, USA}

\author{C. Cariker}%
\affiliation{Department of Physics, University of Central Florida, Orlando, Florida 32816, USA}

\author{M. Shaikh}%
\affiliation{Department of Physics, University of Arizona, Tucson, Arizona 85721, USA}

\author{L. Argenti}%
\email{luca.argenti@ucf.edu}
\affiliation{Department of Physics, University of Central Florida, Orlando, Florida 32816, USA}
\affiliation{CREOL, University of Central Florida, Orlando, Florida 32816, USA}

\author{A. Sandhu}%
\email{asandhu@arizona.edu}
\affiliation{Department of Physics, University of Arizona, Tucson, Arizona 85721, USA}
\affiliation{College of Optical Sciences, University of Arizona, Tucson, Arizona 85721, USA}

\date{\today}%


\begin{abstract}

Tunable attosecond transient absorption spectroscopy is an ideal tool for studying and manipulating autoionization dynamics in the continuum. We investigate near-resonant two-photon couplings between the bright \fp and dark \ff autoionizing states of argon that lead to Autler-Townes like interactions, forming entangled light-matter states, or polaritons. We observe that one-photon couplings with intermediate dark states play an important role in this interaction, leading to the formation of multiple polaritonic branches whose energies exhibit avoided crossings as a function of the dressing-laser frequency. Our experimental measurements and theoretical essential-state simulations show good agreement and reveal how the delay, frequency, and intensity of the dressing pulse govern the properties of autoionizing polariton multiplets. These results demonstrate new pathways for quantum control of autoionizing states with optical fields.

\end{abstract}

\maketitle


\section{Introduction}\label{sec:intro}


Attosecond spectroscopy is a powerful and versatile technique to investigate time-resolved electron dynamics in atomic and molecular systems~\mbox{\cite{Krausz2009AttosecondPhysics,Goulielmakis2010Real-timeMotion}}. In particular, the study of autoionizing states (AISs) is essential for understanding electron-core interactions. It is well known that AISs exhibit asymmetric line shapes in their photoabsorption spectra due to the presence of discrete bound states that are coupled to different continua. Electrons excited to these states autoionize on timescales of tens to hundreds of femtoseconds due to configuration interactions with the core. The absorption features of AISs are described by the familiar Fano profile~\mbox{\cite{Fano1961EffectsShifts,Fano1965LineGases}}.  These line shapes usually contain a region of higher transparency compared to the continuum background which results from the destructive interference between the transition amplitudes of the direct ionization to the continuum and the indirect ionization to this same continuum through the AIS.

Attosecond transient absorption spectroscopy (ATAS) has been extensively used to study AISs~\mbox{\cite{Wang2010argon,Ott2013LorentzLine,Li2015InvestigationXenon,Kaldun2016ObservingDomain,Liao2017Oxygen,Hutten2018UltrafastKrypton}}. In particular, strong-field ATAS~\mbox{\cite{Wu2016TheoryAbsorption}}, which employs  extreme-ultraviolet (XUV) pulses to coherently launch a dipole excitation in an atomic or molecular system, and time-delayed infrared (IR) pulses to control and/or probe its evolution, offers exciting opportunities for control of metastable states of matter~\cite{Lambropoulos1981AutoionizingFields,Kim1982Laser-IntensityAutoionization,Chen2012Helium}. 

In our recent work~\cite{Harkema2021AutoionizingIonization}, we used ATAS to investigate one-photon couplings between a bright AIS and nearby dark AISs via laser dressing. In the near-resonant condition, akin to Autler-Townes phenomena~\mbox{\cite{ZHLoh2008ATS,Pfeiffer2012ATS, Chini2013ATS,Wu2013ATS,Chini2014ATS,Argenti2015ATS,Kobayashi2017ATS,Harkema2018Noncommensurate}}, the interaction between the bright AIS and the autoionizing light-induced state (ALIS) of the dark state leads to the formation of two entangled light-matter states, known as autoionizing polaritons (AIPs). The AIP can decay either through autoionization (AI) or via radiative ionization (RI). We demonstrated that these decay pathways can add coherently, and under specific conditions this interference can be destructive, leading to stabilization against ionization~\mbox{\cite{Lambropoulos1981AutoionizingFields, Harkema2021AutoionizingIonization}}.

In this work, we extend our study to investigate the role of two-photon couplings in the formation of AIPs and explore multi-polariton formation at higher IR intensities. We demonstrate that tunable-dressing-field ATAS is an ideal tool for resolving and controlling polaritonic interactions. Contrary to traditional ATAS, where the XUV and IR frequencies are commensurate, our tunable approach employs an independently adjustable IR probe frequency ($\omega_{\mathrm{IR}}$) to resonantly drive or detune the light-induced couplings between different excited states~\mbox{\cite{Harkema2018Noncommensurate, Harkema2021AutoionizingIonization}}. Specifically, the frequency tunability of the dressing IR field allows us to manipulate the interactions between the \fp bright AIS and several ALISs of the neighboring dark states ($3s^{-1}3d$, $3s^{-1}5s$, and $3s^{-1}4f$) in argon, thus providing control over the formation and evolution of multiple AIPs. We observe that even under resonant conditions for the $3s^{-1}4p\,{-}\,3s^{-1}4f$ two-photon coupling, the intermediate one-photon couplings with other dark states play an important role, leading to the formation of up to four polaritonic branches at higher intensities. We systematically vary the delay, frequency, and intensity of the IR pulse to explore the parameter space and obtain excellent agreement with \mbox{\emph{ab initio}} theory. By selectively incorporating states into the simulations, we are able to identify contributions of each polaritonic interaction, and observe the avoided crossings between various branches.


\begin{figure*}[t]
\centering 
\includegraphics[width=\textwidth]{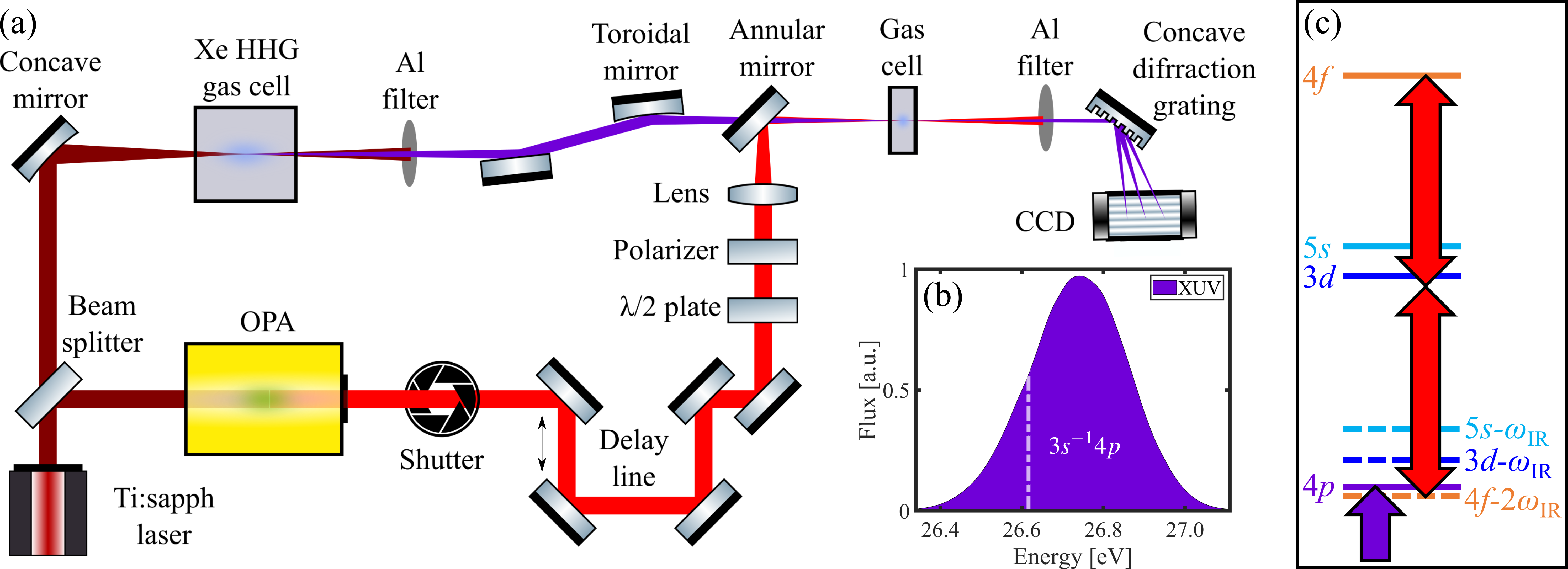}
    \caption{(a) Experimental setup for tunable-dressing-field ATAS displaying the XUV and IR arms. An OPA is used to generate tunable IR pulses for investigation of one-photon and two-photon couplings leading to strong near-resonant interactions between AISs and ALISs. (b) Normalized XUV spectrum corresponding to the 17\textsuperscript{th} harmonic, relative the energy of the bright \fp{} AIS. (c) Relevant AISs and ALISs in argon. The red double arrows show the IR induced $3s^{-1}4p\,{-}\,3s^{-1}4f$ two-photon coupling.}
    \label{fig:exp_setup}
\end{figure*}



This paper is organized as follows. Section~\ref{sec:exp_setup} describes the experimental setup and techniques for tunable ATAS. In Sec.~\ref{sec:theory}, we introduce a theoretical framework for modeling AIP dynamics.  In Sec.~\ref{sec:results}, we present our experimental photoabsorption results and compare them with theoretical essential-state simulations. Finally, in Sec.~\ref{sec:conclusions}, we summarize our work and provide perspectives for future research.


\section{Experimental setup}\label{sec:exp_setup}

We use tunable ATAS to resolve and manipulate AIP dynamics involving multiple AISs. Our pump-probe experimental scheme, shown in Fig.~\ref{fig:exp_setup}(a), employs XUV attosecond pulse trains (APT) to resonantly excite the \fp{} AIS in argon, which can be subsequently probed and controlled through couplings induced by tunable IR pulses. Initially, 1.8~mJ, 40~fs near-infrared (NIR) pulses are generated by a 1 kHz Ti:sapphire laser amplifier with a central wavelength of 790~nm. The beam is then divided into two arms with a 50/50 beam splitter. The NIR pulse in one of the arms is focused into a semi-infinite xenon filled gas cell (10~Torr backing pressure) with a 50~cm focal length curved mirror to produce XUV APT through high harmonic generation (HHG)~\cite{Corkum1993PlasmaIonization,Antoine1996AttosecondHarmonics,Paul2001ObservationGeneration}. The resulting XUV spectrum consists predominantly of the 13$^\mathrm{th}$ to 19$^\mathrm{th}$ odd harmonics, with phase-matching conditions chosen to maximize the XUV flux in the 17$^\mathrm{th}$ harmonic [Fig.~\ref{fig:exp_setup}(b)], which is resonant with the bright \fp{} AIS. The driving NIR pulse is filtered out with a 200~nm aluminum filter and the XUV APT are focused with a toroidal mirror into a second gas cell for the transient absorption study. This gas cell is 3~mm thick and contains argon at a backing pressure of 11~Torr.

The other half of the NIR pulse is routed to an optical parametric amplifier (OPA), which is capable of generating tunable IR pulses in the 0.73--1.03~eV range with an approximate conversion efficiency of 20\%. A half-wave plate and a polarizer are used to control the intensity of the IR probe pulse and a 100~cm lens is used to focus the IR pulse into the argon-filled transient absorption gas cell. A translation stage provides control over the relative time-delay between the XUV APT and the IR pulses. Both beams are combined using an annular mirror such that they propagate collinearly to the argon gas cell, where they are spatio-temporally overlapped. After the interaction, the co-propagating IR beam is filtered out with a second 200-nm-thick aluminum filter and the XUV APT spectrum is diffracted and refocused by a reflective variable line-space grating onto a CCD camera. The resolution of our spectrometer is 15~meV at 24.0~eV. The experimental IR photon energies reported in this manuscript have an uncertainty of $\pm$0.01~eV and agree with those in the essential states model within 20~meV.

We characterize our results in terms of the absolute optical density (OD), calculated as: \mbox{$\mathrm{OD}\,{=}\,{-}\log_{10}\left( I_\mathrm{XUV+IR} / I_{\mathrm{ref}} \right)$}, where $I_{\mathrm{ref}}$ is the reference XUV spectrum through the empty gas cell \mbox{[Fig.~\ref{fig:exp_setup}(b)]} and $I_\mathrm{XUV+IR}$ is the transmitted IR-dressed XUV spectrum through the argon-filled gas cell. In this method, compared to the background OD associated with the continuum absorption, a dip in the OD would indicate a Fano window resonance. In general, one can observe various Fano profiles associated with bright AISs or AIP multiplets resulting from the interaction of a bright AIS with different ALIS counterparts. Positive time-delays \mbox{($\tau\,{=}\,t_{\mathrm{XUV}}{-}t_{\mathrm{IR}}$)} mean the IR dressing field precedes the XUV APT. 

This paper focuses on near-resonant two-photon couplings between laser-dressed AISs, \mbox{$\sum_a |a\rangle\,{\otimes}\,|N_a\gamma_{\textsc{IR}}\rangle$}, where \mbox{$|a\rangle\,{\otimes}\,|N_a\gamma_{\textsc{IR}}\rangle$} (or \mbox{$a\,{\otimes}\,N_a\gamma_{\textsc{IR}}$}, for brevity) represents a component in which the atom is in AIS $|a\rangle$ and the radiation field has $N_a$ IR photons, as well as the properties of the polaritonic states resulting from such interaction. As shown in Fig.~\ref{fig:exp_setup}(c), we tuned the IR photon energy so that the \fflis{} ALIS is near-resonant with the $3s^{-1}4p\,{\otimes}\,N\gamma_{\textsc{IR}}$ AIS. In addition, we systematically vary the IR peak intensity between 20 and 200~GW/cm$^{2}$ to study the intensity dependence of the polaritonic structures. 


\begin{figure}[b]
\centering \includegraphics[width=\columnwidth]{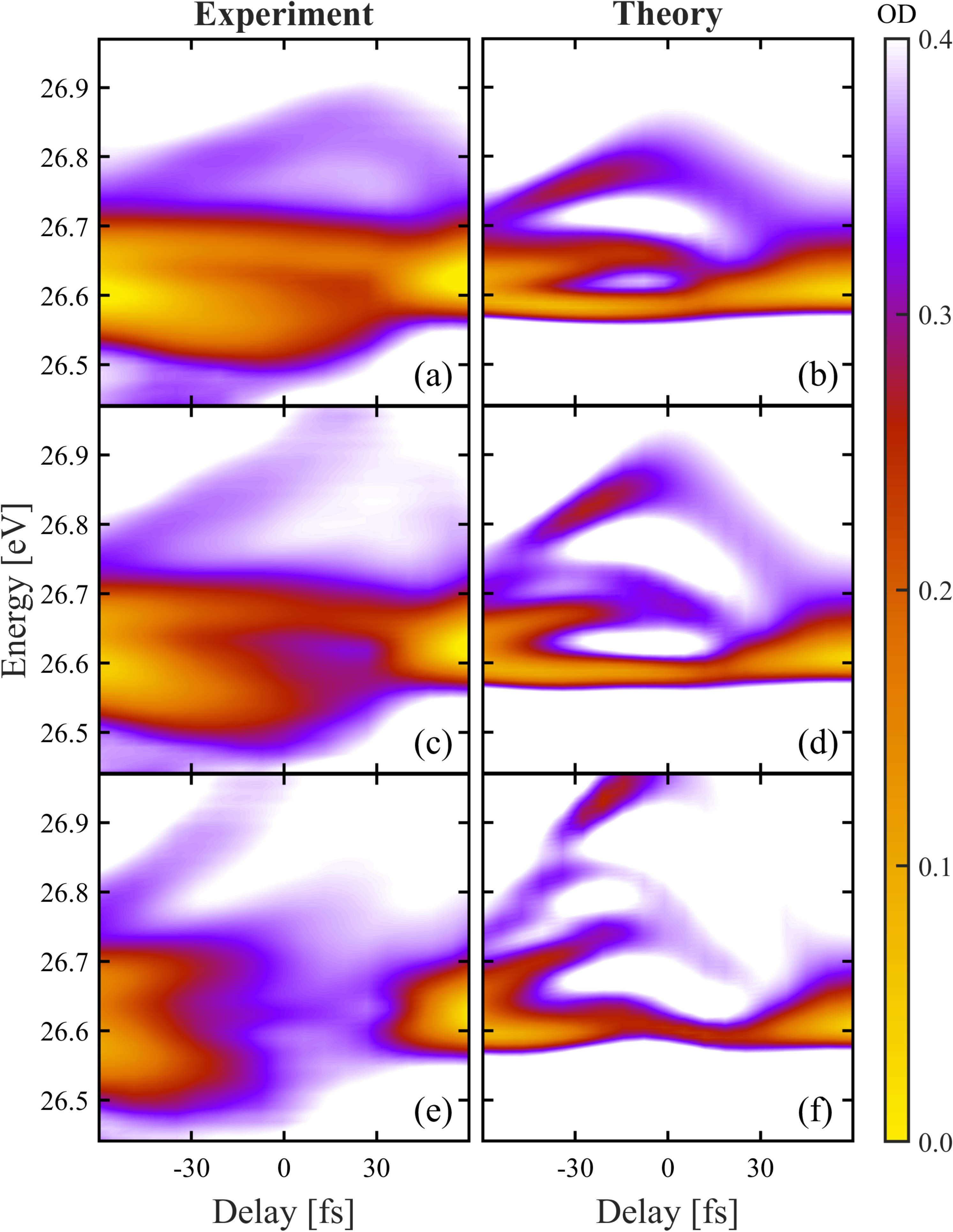}
    \caption{(a), (c), (e)~Experimental and (b), (d), (f)~theoretical XUV photoabsorption spectra in the vicinity of the bright AIS as a function of time-delay. The \fp{} state is near-resonant with the \ff state through a two-photon transition. The dressing field frequency is $\omega_\mathrm{IR}\,{=}\,0.85$~eV for the experiment and $\omega_\mathrm{IR}\,{=}\,0.83$~eV for theory. Different rows illustrate various IR intensities, increasing from top to bottom, where \mbox{(a) 62 GW/cm$^2$}, \mbox{(b) 50 GW/cm$^2$}, \mbox{(c) 110 GW/cm$^2$}, \mbox{(d) 100 GW/cm$^2$}, \mbox{(e) 205 GW/cm$^2$}, and \mbox{(f) 200 GW/cm$^2$}. The color map represents the OD, where lower values indicate decrease in background continuum absorption due to the presence of the \fp AIS or AIP resonances.}
    \label{fig:delay_comparison}
\end{figure}



\section{Theoretical methods}\label{sec:theory}

The experimental photoabsorption spectra are interpreted by using \mbox{\emph{ab initio}} theoretical calculations conducted with the NewStock program~\cite{Carette2013,Marante2017,Chew2018}. The multi-configuration states of the parent ions are computed with the ATSP2K atomic-structure package~\cite{Froese2007}, using the multi-configurational Hartree-Fock (MCHF) method, optimized for the $3s^{-1}$ and $3p^{-1}$ parent ions. The resulting MCHF localized active orbitals are used to form a basis of localized configurations for the neutral system (localized channel). A finite set of ionic states are augmented by a set of diffuse spherical orbitals, complementary to the MCHF active orbitals, given by products of spherical harmonics and radial B-splines (partial-wave channels)~\cite{Argenti2016}. The set of partial-wave channels together with the localized channel form a so-called close-coupling configuration space that is able to accurately reproduce the asymptotic multi-channel character as well as the short-range correlated character of bound states, Rydberg states, and the resonant electronic continuum, within a given quantization box, which in the present work has a radius \mbox{$R_{\mathrm{box}}\,{=}\,500$~Bohr-radii}. The Hamiltonian and dipole matrix elements between close-coupling states are computed. As in any explicit simulation of a continuum system within a confined space, it is necessary to prevent unphysical reflections of the photoelectron wavepacket at the quantization-box boundary. In the present work, this is achieved by adding, to the field-free Hamiltonian $H_0$, a smooth monoelectronic complex absorption potential (CAP), \mbox{$V_{\mathrm{cap}}(\vec{r}_1,\vec{r}_2,\ldots,\vec{r}_n)\,{=}\,\sum_j v(r_j)$}, where \mbox{$v(r)\,{=}\,{-}i\,c\,\theta(r{-}R_{\mathrm{cap}})\,(r{-}R_{\mathrm{cap}})^2$}, $c$ is a real positive number, and $\theta(x)$ is the Heaviside step function. This potential, which starts ${\sim}$100~Bohr-radii before the box boundary, makes any outgoing wave function vanish before it can be reflected. While this approach dissipates the outermost part of the wave function, it preserves the internal portion, which is the only one relevant to reproduce the attosecond transient absorption spectrum of the atom. Indeed, the optical response of the atom in the XUV spectral region is due to the dipole beating between the excited component of the wave function and the ground state, which is highly localized near the origin.
The CAP enforces outgoing boundary conditions on the eigenfunctions of the Hamiltonian of the confined system (Siegert states~\cite{Siegert1939,Tolstikhin1997,Tolstikhin2006}), $H_{\mathrm{cap}}\,{=}\,H_0{+}V_{\mathrm{cap}}$, above the ionization threshold, which acquire finite lifetimes and hence a complex energies. The AISs, in particular, emerge as isolated complex eigenvalues, and hence can be selectively included or excluded in the simulation of the laser-driven dynamics of the system, thus revealing the role of individual resonances. In the case of optical observables, the diagonalization of the field-free Hamiltonian allows us also to restrict the simulation space to few hundreds essential states, whose energy lies below a prescribed cutoff, thus reducing the cost of simulation by several orders of magnitude, compared with a simulation in the complete basis, without compromising the convergence of the observables of interest here.

To compute the evolution of the system in the presence of external radiation fields, we numerically solve the time-dependent Schr\"{o}dinger equation (TDSE) in a succession of time steps, \mbox{$|\Psi(t+dt)\rangle = U(t+dt,t)|\Psi(t)\rangle$}, using a second-order split-exponential propagator \mbox{$U(t+dt,t)$}, where
\begin{equation}
\begin{split}
U(t+dt,t) = e^{-i H_{\mathrm{cap}}\frac{dt}{2}}\, e^{  -i\,H_I\left(t+\frac{dt}{2}\right)\,dt}\,e^{-iH_{\mathrm{cap}} \frac{dt}{2}}. 
 \end{split}
 \end{equation}
 
In the velocity gauge and in the dipole approximation, \mbox{$H_I(t)\,{=}\,\alpha \vec{A}(t){\cdot} \vec{P}(\vec{r})$}, where $\vec{A}(t)$ is the field vector potential of both the pump and probe pulses, \mbox{$\vec{A}(t)\,{=}\,\vec{A}_{\mathrm{XUV}}(t-\tau){+}\vec{A}_{\mathrm{IR}}(t)$}, $P(\vec{r})$ is the electronic canonical momentum, and $\alpha$ is the fine-structure constant. The exponential of the dipole matrix is computed exactly using its diagonal representation in the essential Siegert-states basis. Therefore, each time step is extremely fast, since it merely requires two matrix-vector operations, which scale quadratically with the size of the Siegert-space basis. For optically thin samples, the absorption at a given frequency is related to the Fourier transform of the expectation value of the dipole operator, given by $\sigma(\omega)\,{=}\,{-}4\pi/\omega\Im\lbrace{\tilde{P}(\omega)}/{\tilde{A}(\omega)}\rbrace$.

As discussed in the following section, we have conducted essential state simulations where we retain only a specific set of states to quantify the relative contribution of various light-induced couplings between specific bright and dark AISs. 


\section{Results and Discussion}\label{sec:results}


The focus of our investigations is the near-resonant interaction between the bright \fp and dark \ff AISs due to the two-photon coupling induced by the IR field. Figure~\ref{fig:delay_comparison} shows a comparison between the experimental (left) and theoretical (right) photoabsorption spectrograms, at three different IR intensities (rows), in the vicinity of  the bright \fp{} AIS. At longer time-delays, we solely observe the \fp{} window resonance. In contrast, when the XUV and IR pulses overlap near zero delay, multiple IR-field-induced polaritonic structures emerge in the continuum. Although we tuned the IR photon energy for a resonant two-photon 4\emph{p}--4\emph{f} coupling ($\omega_\mathrm{IR}\,{\sim}\,0.84$~eV), the observed multiplicity of features points to the presence of additional interactions. 

At lower IR intensities ($\sim$50~GW/cm$^2$), the experimental and theoretical photoabsorption data in Figs.~\ref{fig:delay_comparison}(a) and \ref{fig:delay_comparison}(b) respectively, show the \fp line shape splits into three dominant AIP features. As the intensity is increased to $\sim$100~GW/cm$^2$ and above [Figs.~\ref{fig:delay_comparison}(c--f)], the splitting between different AIP branches increases substantially, and the top branches also exhibit discontinuity, as seen around 26.8~eV at $\tau\,{\approx}\,25$~fs. In general, the agreement between the experiment and theory is very good. At high intensities, we clearly observe four AIP branches in both experiment and theory. These observations are in contrast to the expectation of two AIP branches stemming from the resonant interaction between the $3s^{-1}4p$ and the $3s^{-1}4f$ states.

\begin{figure}[t]
\centering \includegraphics[width=\columnwidth]{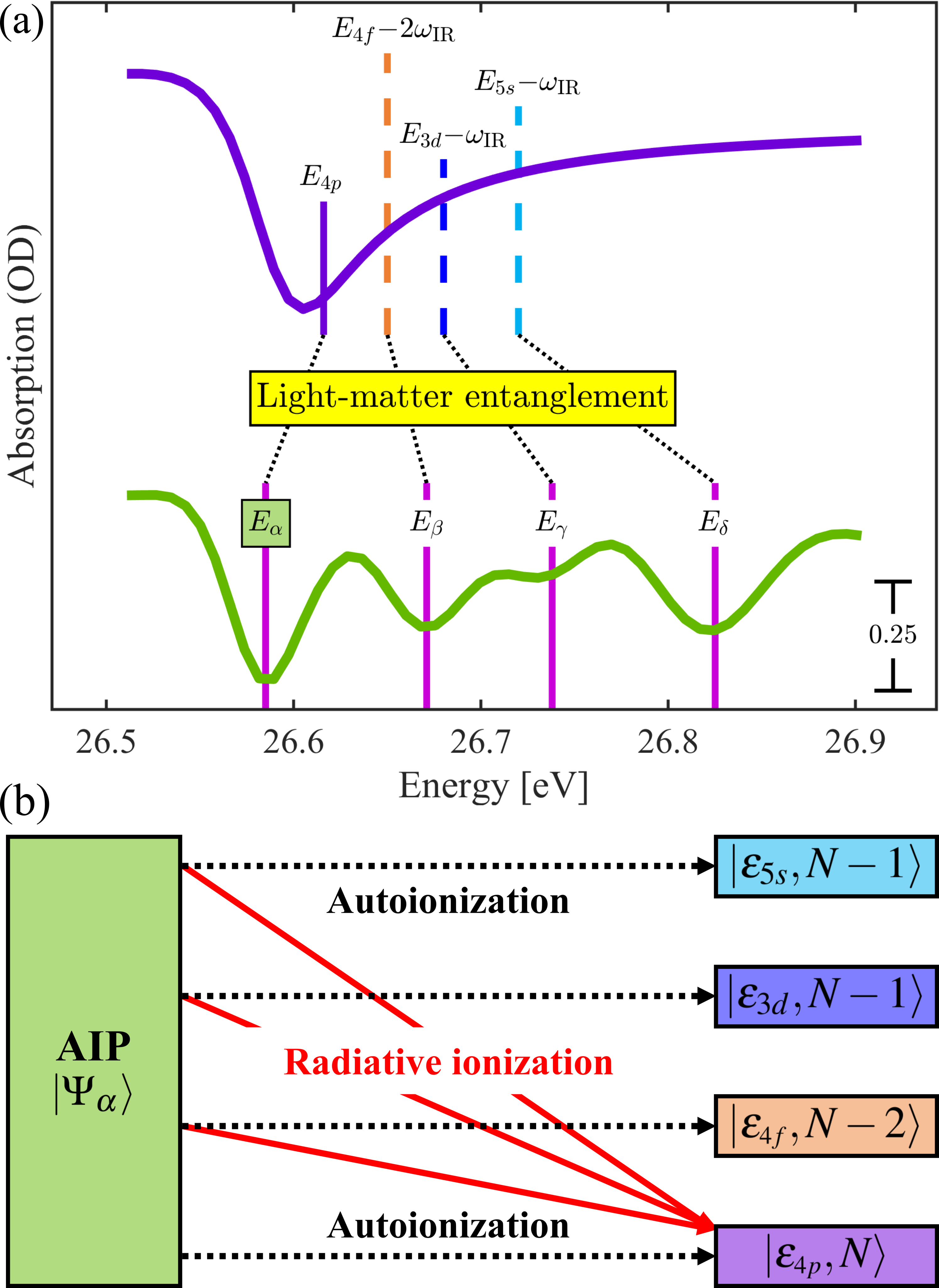}
    \caption{(a) Diagrammatic representation of the interacting bright (4$p$) and nearby dark (3$d$, 5$s$, 4$f$) $3s^{-1}n\ell$ AISs in argon. The resultant polaritonic multiplets represent entangled light-matter states. The simulated photoabsorption spectrum with XUV alone (purple) is a window resonance and the IR-dressed spectrum (green) clearly shows four AIP branches. (b) Each of these AIPs can decay to the continuum states through AI and RI pathways, which interference constructively or destructively, depending on the IR laser parameters.} 
    \label{fig:energy_diagram}
\end{figure}

The light-induced redistribution of the \fp line strength is a direct consequence of the renormalization of the AIS and ALIS energies. Despite being rather detuned from the one-photon resonance condition [Fig.~\ref{fig:exp_setup}(c)], the \tdlis{} and \fslis{} ALISs can also contribute to the interaction. In fact, as shown in Fig.~\ref{fig:energy_diagram}, a basis set involving the $3s^{-1}4p\,{\otimes}\,N\gamma_{\mathrm{IR}}$~AIS and the \tdlis, \fslis, and \fflis{} ALISs, qualitatively explains the observations. The interaction between these four states gives rise to four entangled light-matter states, which matches the number of polaritonic branches seen in our data in Figs.~\ref{fig:delay_comparison}(c-f).


In a recent work~\cite{Harkema2021AutoionizingIonization}, we employed an extended Jaynes-Cummings (J-C) model to account for the formation and decay of AIPs~\mbox{\cite{JCM1963,greentree2013fiftyJCM}}. This allowed us to understand the coherent interferences between AI and RI decay pathways, paving the path for optical control of the metastable states. Our analysis considered a situation with two laser-coupled states, that lead to the formation of two distinct polaritonic branches. In the current work, we highlight how there are many more possibilities for interfering paths, and how both one and two-photon couplings are active. 
Figure \ref{fig:energy_diagram}(a) shows the energy diagram of the ALISs near the \fp state in argon (26.62~eV). Initially, an XUV photon promotes the electron from the ground state to the bright AIS. The calculated absorption profile of the \fp{} state, shown in the purple curve, represents a single Fano window resonance.  Subsequently, a time-delayed tunable IR pulse couples to the nearby dark AISs, namely $3s^{-1}3d$ (27.51~eV), $3s^{-1}5s$ (27.55~eV), and $3s^{-1}4f$ ($\sim$28.31~eV). Tuning the IR photon energy allows us to manipulate the relative position of different ALISs, such as \tdlis, \fslis, and \fflis{}, with respect to the \fp{} state. In particular, for this study, we bring the 4$p$ and 4$f$ close to a two-photon resonance. However, due to their relatively close spacing, the bright AIS and the ALISs form a strongly interacting basis, leading to the formation of AIPs multiplets with renormalized energy structure ($E_\alpha,E_\beta,E_\gamma,E_\delta$), as shown in Fig.~\ref{fig:energy_diagram}(a). The calculated photoabsorption profile of XUV in the presence of IR couplings (green curve), which is a lineout of Fig.~\ref{fig:delay_comparison}(d) at $\tau\,{=}\,{-}30$~fs, clearly shows four polaritonic features in agreement with our high intensity experimental observations in Figs.~\ref{fig:delay_comparison}(c) and \ref{fig:delay_comparison}(e).

Fig. \ref{fig:energy_diagram}(b) illustrates the decay paths of a particular AIP state, $\ket{\Psi_{\alpha}}$, which results from the mixing of four AISs of different electronic symmetry through one-photon and two-photon couplings. In the extended J-C model, such state can be written as:
\begin{align}
\begin{split}
    \ket{\Psi_\alpha} &= c_p \ket{4p}{\otimes}\ket{N} + c_{d}\ket{3d}{\otimes}\ket{N{-}1}\\
    &+ c_{s}\ket{5s}{\otimes}\ket{N{-}1} + c_{f}\ket{4f}{\otimes}\ket{N{-}2},
\label{eqs:polariton_states}
\end{split}
\end{align}
where $c_\ell$ are complex coefficients. The energy for a given polaritonic branch will be determined by the IR coupling matrix elements and the detunings. The components of the AIP state $\ket{\Psi_{\alpha}}$ can decay directly through configuration interaction to the same symmetry continua, and through radiative couplings to the other continua. For example, if we consider the decay pathways to the $N$-photon $4p$ continuum channel, there will be one direct AI path, and three RI paths, as shown in Fig.~\ref{fig:energy_diagram}(b). The crucial point is that the decay amplitudes of these terms interfere coherently, giving rise to a partial decay rate to the $N$-photon channel that exhibits destructive or constructive interference,
\begin{equation}
    \Gamma_{N,\alpha} = |c_{p} \Gamma^{1/2}_{\mathrm{AI},4p}
    + c_{d} \Gamma^{1/2}_{\mathrm{RI},3d}
    + c_{s} \Gamma^{1/2}_{\mathrm{RI},5s}
    + c_{f} \Gamma^{1/2}_{\mathrm{RI},4f} |^2,
\label{eqs:polariton_widths}
\end{equation}
reflected also in the net decay rate of the polariton, where $\Gamma_{\mathrm{AI},4p}$ is the 4$p$ field-free AI decay rate, while $\Gamma_{\mathrm{RI},3d}$ and $\Gamma_{\mathrm{RI},5s}$ are the one-photon RI rates to the continuum and $\Gamma_{\mathrm{RI},4f}$ is the two-photon RI term. The one-photon (two-photon) rates depend linearly (quadratically) on the IR intensity, and $c_\ell$ depends both on the field strength and detuning, therefore, the IR pulse parameters are important knobs for controlling AIP dynamics. 

\begin{figure}[t]
\centering \includegraphics[width=\columnwidth]{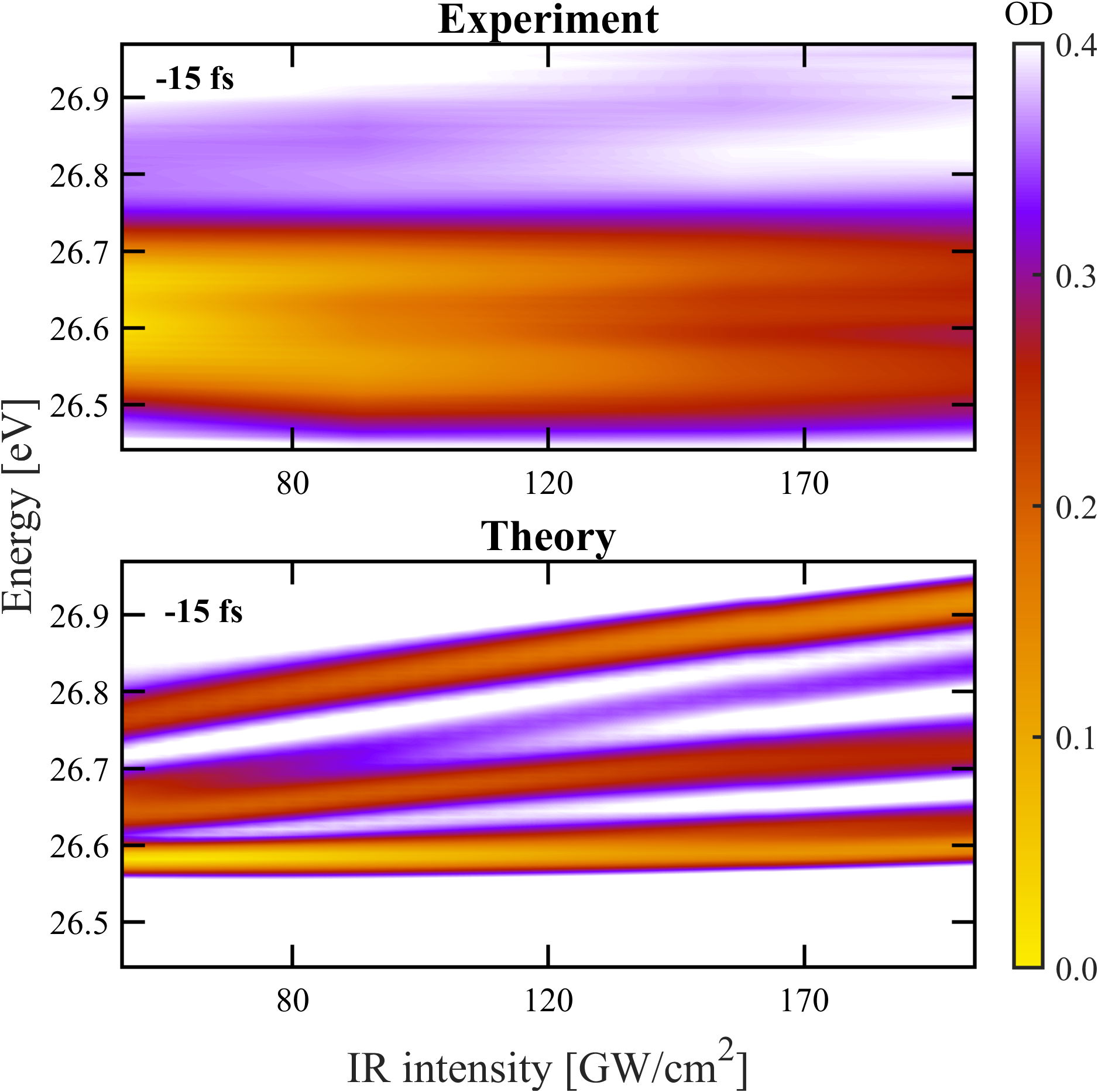}
    \caption{Experimental (top) and theoretically simulated (bottom) IR intensity scan at fixed time-delay \mbox{($\tau\,{=}\,{-}15$~fs)} in the proximity of the \fp AIS, for $\omega_\mathrm{IR}\,{=}\,0.85$~eV and $\omega_\mathrm{IR}\,{=}\,0.83$~eV, respectively. The separation between AIP branches increases with the IR intensity, and their relative strengths and widths are impacted due to coupling effects.}
    \label{fig:single_int_comparison}
\end{figure}

To explore the effects of the dressing field strength over the AIP multiplets, we perform an IR intensity scan keeping the time-delay between the XUV and IR pulses fixed \mbox{($\tau\,{=}\,{-}15$~fs)}, as shown in Fig.~\ref{fig:single_int_comparison}. As previously observed in Fig.~\ref{fig:delay_comparison}, both theory and experiment exhibit four branches as the IR intensity is increased. The AIP splitting continually increases with the field strength in accordance with the expectation. The width of the polaritonic branches also changes with intensity. The complexity of the interactions and the number of interfering terms, however, prevents us from fitting the data with an analytical resonant profile. The differences between Fig.~\ref{fig:single_int_comparison} (a) and (b), in terms of the intensity slope of the branches and relative strength of the different branches, is due to the slight differences between the experimental and theoretical parameters, in addition to the uncertainties in the delay, intensity, and focal volume averaging effect inherent to the experiments.      



\begin{figure*}[t]
\centering \includegraphics[width=\textwidth]{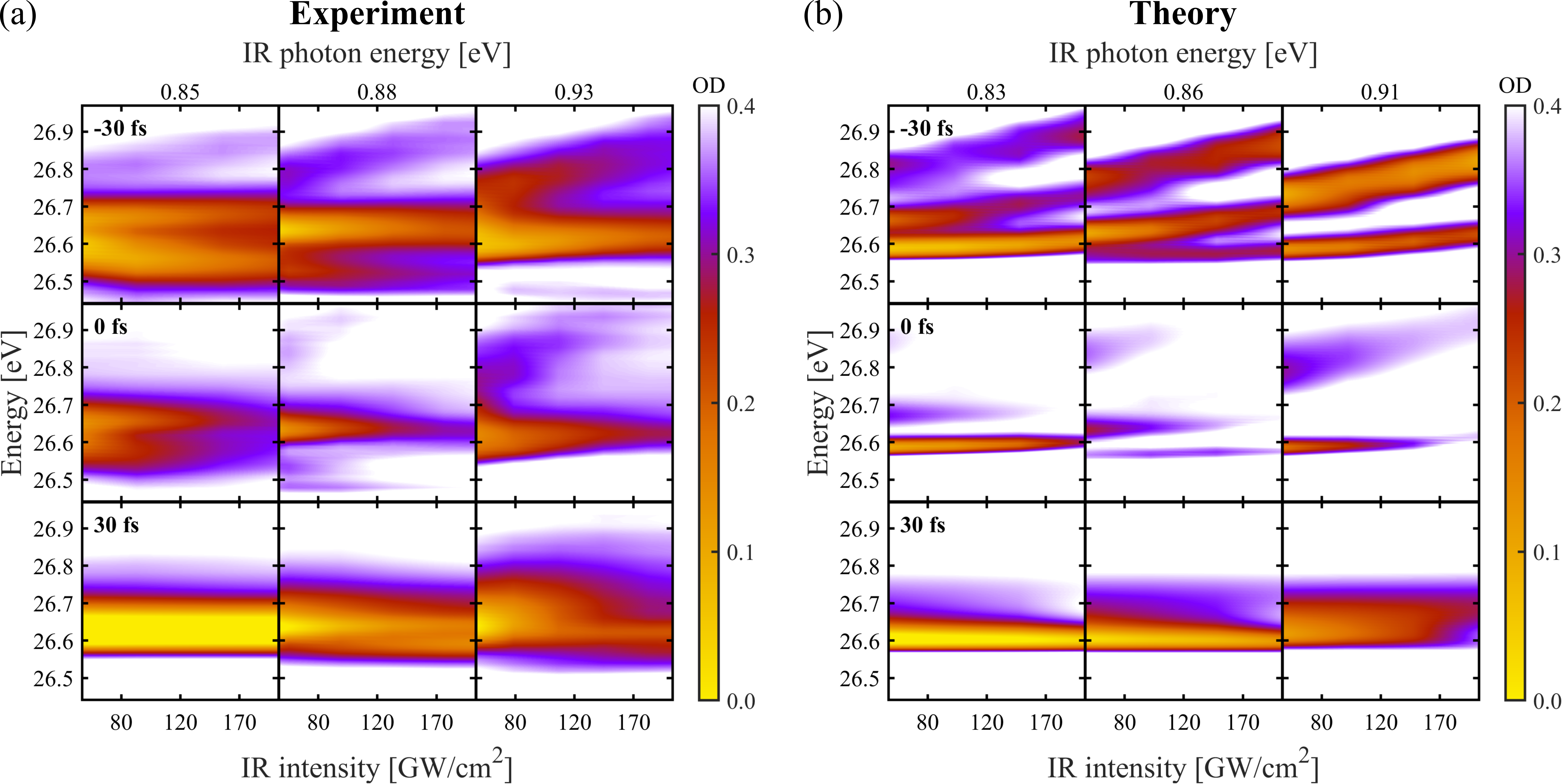}
    \caption{IR intensity dependent XUV photoabsorption spectra in the vicinity of the \fp{} AIS, both for (a)~experiment and (b)~theoretical simulations. The columns indicate different values of $\omega_\mathrm{IR}$ and the rows illustrate fixed time-delays. These plots demonstrate that the frequency and intensity of the IR pulse can be used for optical control of AIPs.}
    \label{fig:mosaic_int_comparison}
\end{figure*}


In Figs.~\ref{fig:mosaic_int_comparison}(a) and \ref{fig:mosaic_int_comparison}(b), we expand the analysis of the intensity dependence of polaritonic multiplets  to include additional IR delays and detuning values. Here, each row corresponds to a specific time-delay and each column to a different IR photon energy. Clearly, the number and the nature of splittings is highly dependent on these parameters. Apart from the ${\sim}$0.02~eV offset in the IR frequencies between experiment and theory panels, the agreement between the two is notable and the \mbox{\emph{ab initio}} theory captures all the details from our systematic experimental study.  

For computational time considerations, the theoretical simulations shown in Fig.~\ref{fig:mosaic_int_comparison}(b) have lower resolution intensity steps than the one shown in Fig.~\ref{fig:single_int_comparison}. As a result, some of the finer details of AIP splitting are not visible. Nevertheless, the trends with detuning, time-delay, and peak intensity are evident. At positive time-delays \mbox{($\tau\,{=}\,30$ fs)}, both Figs.~\ref{fig:mosaic_int_comparison}(a) and \ref{fig:mosaic_int_comparison}(b) exhibit fewer features and smaller splittings, which is expected as the IR precedes the XUV APT. In the case of negative time-delays \mbox{($\tau\,{=}\,{-}30$ fs)}, the full IR pulse interacts with the XUV-initiated dipole. Higher IR peak intensities produce larger change in OD, implying an increased redistribution of the \fp{} window resonance into polaritonic branches. Furthermore, Figs.~\ref{fig:mosaic_int_comparison}(a) and \ref{fig:mosaic_int_comparison}(b) show how the IR photon energy tuning can be used to control the AIP structure. At \mbox{$\omega_\mathrm{IR}\,{<}\,0.90$~eV}, when the 4\emph{p}--4\emph{f} coupling is near-resonant, we observe several AIP branches in the XUV spectrum; however, as we move away from resonance with the \ff{} AIS \mbox{($\omega_\mathrm{IR}\,{\ge}\,0.90$~eV)}, we observe two main polaritonic branches corresponding to the interaction of 4$p$ with the closely spaced 3$d$/5$s$ states through one-photon coupling. 

\begin{figure*}[t]
\centering \includegraphics[width=\textwidth]{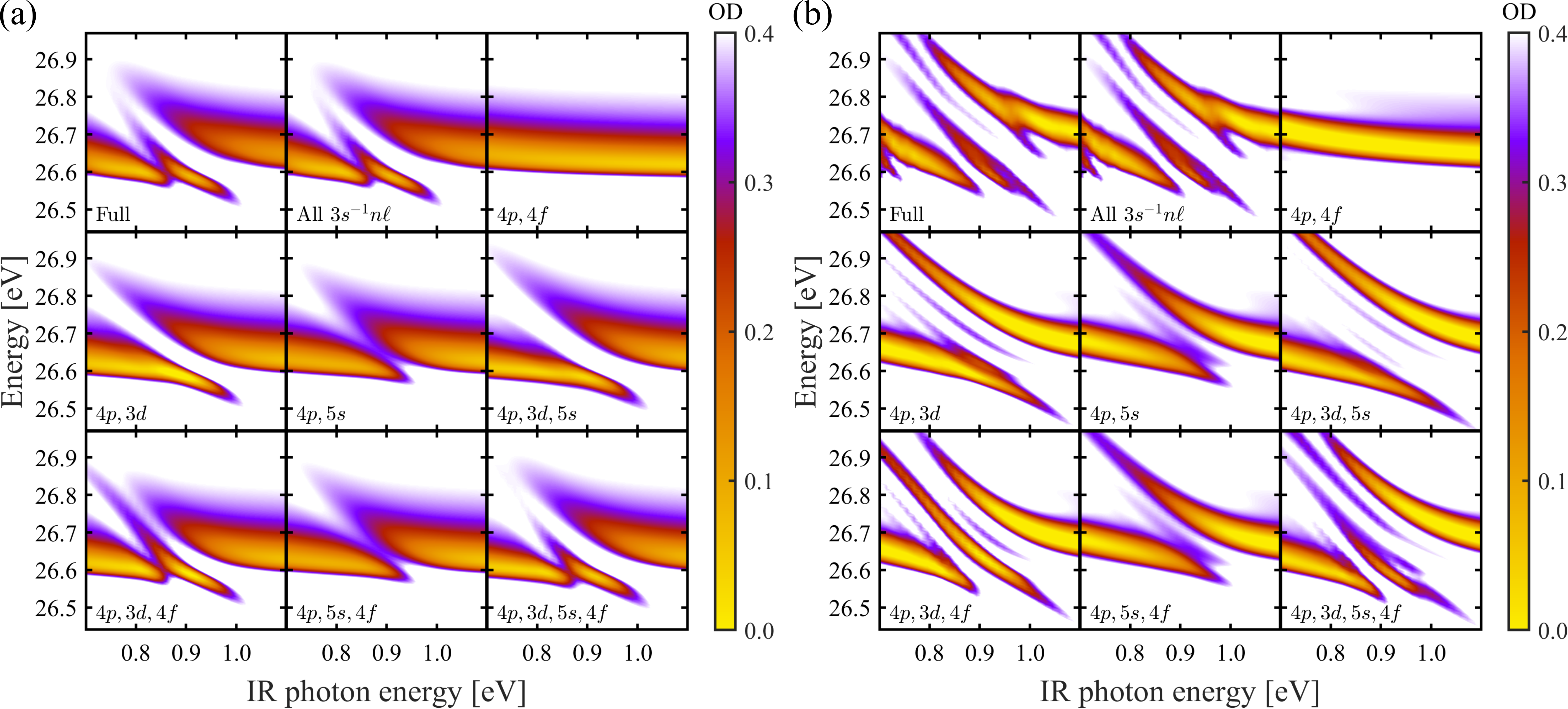}
    \caption{Theoretical dressing field frequency scans at IR intensities of (a) 50 GW/cm$^2$ ($\tau\,{=}\,0$~fs) and (b) 150 GW/cm$^2$ ($\tau\,{=}\,{-}15$~fs). Each panel corresponds to different essential states bases to isolate the individual contribution of different resonances to the photoabsorption spectrum near the \fp AIS.}
    \label{fig:essential_states}
\end{figure*}

Following, we explore the AIP spectrum as a function of the IR photon energy. The goal of this exercise is to identify the contribution of different states to the photoabsorption spectrum using essential-state calculations at different IR intensities. However, in contrast to our previous work \cite{Harkema2021AutoionizingIonization}, here we investigate the contribution of individual resonances not by removing these from the full-basis essential-state simulations, but instead by building a few-level basis composed of only certain states. Figure~\ref{fig:essential_states}(a) shows simulated frequency scans at $\tau\,{=}\,0$ and an IR intensity of 50~GW/cm$^2$. Each sub-panel corresponds to different essential-states. The basis used for each calculation are composed of the ground state, the $3s^{-1}$ and $3p^{-1}$ continua, and the resonances indicated in each panel. For example, the top-right panel (4\emph{p},$\,\,$4\emph{f}) is a simulation computed with the $3s^{-1}4p$ and $3s^{-1}4f$ AISs, along with the aforementioned ground state and continua. By selectively evaluating the contribution of individual AISs in the essential-state simulations, we are able to assign observed AIP features to specific couplings between bright and dark AISs. The full basis (top-left) and the $3s^{-1}n\ell$ (top-center) panels are identical, confirming that the observed dynamics can be fully described by the laser induced coupling of various $3s^{-1}n\ell$ states. 


Next, we investigate the role of intermediate resonances in two-photon AIP formation by comparing the top-right (4\emph{p},$\,\,$4\emph{f}), center-left (4\emph{p},$\,\,$3\emph{d}), and bottom-left (4\emph{p},$\,\,$3\emph{d},$\,\,$4\emph{f}) panels of Fig.~\ref{fig:essential_states}(a). This analysis clearly shows that in the absence of a strong one-photon intermediate resonance, such as $3s^{-1}3d$, the degeneracy between the two-photon \fflis{} ALIS and \fp AIS does not lead to polaritonic splitting. Moreover, a direct comparison with the top-right (4\emph{p},$\,\,$4\emph{f}), center (4\emph{p},$\,\,$5\emph{s}), and bottom-center (4\emph{p},$\,\,$5\emph{s},$\,\,$4\emph{f}) panels confirms the $3s^{-1}5s$ is a weaker intermediate resonance than the $3s^{-1}3d$, as it does not lead to the the formation of the additional avoided crossings in bottom-center panel due to the contribution of the \ff AIS.

To identify the interactions leading to four polaritonic branches, previously discussed in Fig.~\ref{fig:single_int_comparison}, we evaluate the essential state contributions at $\tau\,{=}\,{-}15$~fs and an IR intensity of 150~GW/cm$^2$. These results, shown in  Fig.~\ref{fig:essential_states}(b), exhibit up to four branches with larger AIP splittings. This analysis confirms our understanding that four states, namely the \fp AIS and the \tdlis, \fslis, and \fflis{} ALISs, are essential to match the observation of four polaritonic branches in the experimental photoabsorption. The two-photon 4\emph{p}--4\emph{f} coupling alone cannot produce an observable splitting (top-right), and the role of intermediate one-photon coupled 3$d$ and 5$s$ states is essential to obtain the four features, as in bottom-right panel. 


\section{Concluding remarks}\label{sec:conclusions}

In summary, we employ tunable-dressing-field ATAS to investigate and control atomic polariton autoionization. We show that different pulse parameters of the dressing field; namely, photon energy, pulse intensity, and time-delay, can be adjusted to manipulate the formation and subsequent dynamics of AIP multiplets near the \fp AIS in argon. Furthermore, we observe the \mbox{\emph{ab initio}} theory shows remarkable agreement with the experimental measurements. We investigate the role of the \fflis{} two-photon ALIS in the experimental photoabsorption spectrum. Comparison with simulations indicates that intermediate one-photon resonances are essential to produce the observed polaritonic structure. In particular, we observe the \td{} AIS plays a fundamental role in mediating the 4\emph{p}--4\emph{f} coupling that leads to the splitting of the \fp line. Our systematic experimental work and accurate \mbox{\emph{ab initio}} simulations provide remarkable insights into the light-matter couplings, and offer new opportunities for the optical control of metastable states.


\begin{acknowledgments}
S.Y-P., C.C. and M.S. contributed equally. S.Y-P. and A.S. acknowledge support from the U.S. Department of Energy, Office of Science, Office of Basic Energy Science, under Award No. DE-SC0018251.
M.S. acknowledges support from the NSF grants PHY 1912455 and PHY 1919486. 
L.A. and C.C. acknowledge support from the NSF Theoretical AMO Grants No. 1607588 and No. 1912507 and computer time at the University of Central Florida Advanced Research Computing Center.
\end{acknowledgments}


\bibliography{Refs}

\begin{thebibliography}{36}%
\makeatletter
\providecommand \@ifxundefined [1]{%
 \@ifx{#1\undefined}
}%
\providecommand \@ifnum [1]{%
 \ifnum #1\expandafter \@firstoftwo
 \else \expandafter \@secondoftwo
 \fi
}%
\providecommand \@ifx [1]{%
 \ifx #1\expandafter \@firstoftwo
 \else \expandafter \@secondoftwo
 \fi
}%
\providecommand \natexlab [1]{#1}%
\providecommand \enquote  [1]{``#1''}%
\providecommand \bibnamefont  [1]{#1}%
\providecommand \bibfnamefont [1]{#1}%
\providecommand \citenamefont [1]{#1}%
\providecommand \href@noop [0]{\@secondoftwo}%
\providecommand \href [0]{\begingroup \@sanitize@url \@href}%
\providecommand \@href[1]{\@@startlink{#1}\@@href}%
\providecommand \@@href[1]{\endgroup#1\@@endlink}%
\providecommand \@sanitize@url [0]{\catcode `\\12\catcode `\$12\catcode
  `\&12\catcode `\#12\catcode `\^12\catcode `\_12\catcode `\%12\relax}%
\providecommand \@@startlink[1]{}%
\providecommand \@@endlink[0]{}%
\providecommand \url  [0]{\begingroup\@sanitize@url \@url }%
\providecommand \@url [1]{\endgroup\@href {#1}{\urlprefix }}%
\providecommand \urlprefix  [0]{URL }%
\providecommand \Eprint [0]{\href }%
\providecommand \doibase [0]{https://doi.org/}%
\providecommand \selectlanguage [0]{\@gobble}%
\providecommand \bibinfo  [0]{\@secondoftwo}%
\providecommand \bibfield  [0]{\@secondoftwo}%
\providecommand \translation [1]{[#1]}%
\providecommand \BibitemOpen [0]{}%
\providecommand \bibitemStop [0]{}%
\providecommand \bibitemNoStop [0]{.\EOS\space}%
\providecommand \EOS [0]{\spacefactor3000\relax}%
\providecommand \BibitemShut  [1]{\csname bibitem#1\endcsname}%
\let\auto@bib@innerbib\@empty
\bibitem [{\citenamefont {Krausz}\ and\ \citenamefont
  {Ivanov}(2009)}]{Krausz2009AttosecondPhysics}%
  \BibitemOpen
  \bibfield  {author} {\bibinfo {author} {\bibfnamefont {F.}~\bibnamefont
  {Krausz}}\ and\ \bibinfo {author} {\bibfnamefont {M.}~\bibnamefont
  {Ivanov}},\ }\bibfield  {title} {\bibinfo {title} {{Attosecond physics}},\
  }\href {https://doi.org/10.1103/RevModPhys.81.163} {\bibfield  {journal}
  {\bibinfo  {journal} {Reviews of Modern Physics}\ }\textbf {\bibinfo {volume}
  {81}},\ \bibinfo {pages} {163} (\bibinfo {year} {2009})}\BibitemShut
  {NoStop}%
\bibitem [{\citenamefont {Goulielmakis}\ \emph {et~al.}(2010)\citenamefont
  {Goulielmakis}, \citenamefont {Loh}, \citenamefont {Wirth}, \citenamefont
  {Santra}, \citenamefont {Rohringer}, \citenamefont {Yakovlev}, \citenamefont
  {Zherebtsov}, \citenamefont {Pfeifer}, \citenamefont {Azzeer}, \citenamefont
  {Kling}, \citenamefont {Leone},\ and\ \citenamefont
  {Krausz}}]{Goulielmakis2010Real-timeMotion}%
  \BibitemOpen
  \bibfield  {author} {\bibinfo {author} {\bibfnamefont {E.}~\bibnamefont
  {Goulielmakis}}, \bibinfo {author} {\bibfnamefont {Z.~H.}\ \bibnamefont
  {Loh}}, \bibinfo {author} {\bibfnamefont {A.}~\bibnamefont {Wirth}}, \bibinfo
  {author} {\bibfnamefont {R.}~\bibnamefont {Santra}}, \bibinfo {author}
  {\bibfnamefont {N.}~\bibnamefont {Rohringer}}, \bibinfo {author}
  {\bibfnamefont {V.~S.}\ \bibnamefont {Yakovlev}}, \bibinfo {author}
  {\bibfnamefont {S.}~\bibnamefont {Zherebtsov}}, \bibinfo {author}
  {\bibfnamefont {T.}~\bibnamefont {Pfeifer}}, \bibinfo {author} {\bibfnamefont
  {A.~M.}\ \bibnamefont {Azzeer}}, \bibinfo {author} {\bibfnamefont {M.~F.}\
  \bibnamefont {Kling}}, \bibinfo {author} {\bibfnamefont {S.~R.}\ \bibnamefont
  {Leone}},\ and\ \bibinfo {author} {\bibfnamefont {F.}~\bibnamefont
  {Krausz}},\ }\bibfield  {title} {\bibinfo {title} {{Real-time observation of
  valence electron motion}},\ }\href {https://doi.org/10.1038/nature09212}
  {\bibfield  {journal} {\bibinfo  {journal} {Nature}\ }\textbf {\bibinfo
  {volume} {466}},\ \bibinfo {pages} {739} (\bibinfo {year}
  {2010})}\BibitemShut {NoStop}%
\bibitem [{\citenamefont {Fano}(1961)}]{Fano1961EffectsShifts}%
  \BibitemOpen
  \bibfield  {author} {\bibinfo {author} {\bibfnamefont {U.}~\bibnamefont
  {Fano}},\ }\bibfield  {title} {\bibinfo {title} {Effects of configuration
  interaction on intensities and phase shifts},\ }\href
  {https://doi.org/10.1103/PhysRev.124.1866} {\bibfield  {journal} {\bibinfo
  {journal} {Phys. Rev.}\ }\textbf {\bibinfo {volume} {124}},\ \bibinfo {pages}
  {1866} (\bibinfo {year} {1961})}\BibitemShut {NoStop}%
\bibitem [{\citenamefont {Fano}\ and\ \citenamefont
  {Cooper}(1965)}]{Fano1965LineGases}%
  \BibitemOpen
  \bibfield  {author} {\bibinfo {author} {\bibfnamefont {U.}~\bibnamefont
  {Fano}}\ and\ \bibinfo {author} {\bibfnamefont {J.~W.}\ \bibnamefont
  {Cooper}},\ }\bibfield  {title} {\bibinfo {title} {Line profiles in the
  far-uv absorption spectra of the rare gases},\ }\href
  {https://doi.org/10.1103/PhysRev.137.A1364} {\bibfield  {journal} {\bibinfo
  {journal} {Phys. Rev.}\ }\textbf {\bibinfo {volume} {137}},\ \bibinfo {pages}
  {A1364} (\bibinfo {year} {1965})}\BibitemShut {NoStop}%
\bibitem [{\citenamefont {Wang}\ \emph {et~al.}(2010)\citenamefont {Wang},
  \citenamefont {Chini}, \citenamefont {Chen}, \citenamefont {Zhang},
  \citenamefont {He}, \citenamefont {Cheng}, \citenamefont {Wu}, \citenamefont
  {Thumm},\ and\ \citenamefont {Chang}}]{Wang2010argon}%
  \BibitemOpen
  \bibfield  {author} {\bibinfo {author} {\bibfnamefont {H.}~\bibnamefont
  {Wang}}, \bibinfo {author} {\bibfnamefont {M.}~\bibnamefont {Chini}},
  \bibinfo {author} {\bibfnamefont {S.}~\bibnamefont {Chen}}, \bibinfo {author}
  {\bibfnamefont {C.-H.}\ \bibnamefont {Zhang}}, \bibinfo {author}
  {\bibfnamefont {F.}~\bibnamefont {He}}, \bibinfo {author} {\bibfnamefont
  {Y.}~\bibnamefont {Cheng}}, \bibinfo {author} {\bibfnamefont
  {Y.}~\bibnamefont {Wu}}, \bibinfo {author} {\bibfnamefont {U.}~\bibnamefont
  {Thumm}},\ and\ \bibinfo {author} {\bibfnamefont {Z.}~\bibnamefont {Chang}},\
  }\bibfield  {title} {\bibinfo {title} {Attosecond time-resolved
  autoionization of argon},\ }\href
  {https://doi.org/10.1103/PhysRevLett.105.143002} {\bibfield  {journal}
  {\bibinfo  {journal} {Phys. Rev. Lett.}\ }\textbf {\bibinfo {volume} {105}},\
  \bibinfo {pages} {143002} (\bibinfo {year} {2010})}\BibitemShut {NoStop}%
\bibitem [{\citenamefont {Ott}\ \emph {et~al.}(2013)\citenamefont {Ott},
  \citenamefont {Kaldun}, \citenamefont {Raith}, \citenamefont {Meyer},
  \citenamefont {Laux}, \citenamefont {Evers}, \citenamefont {Keitel},
  \citenamefont {Greene},\ and\ \citenamefont {Pfeifer}}]{Ott2013LorentzLine}%
  \BibitemOpen
  \bibfield  {author} {\bibinfo {author} {\bibfnamefont {C.}~\bibnamefont
  {Ott}}, \bibinfo {author} {\bibfnamefont {A.}~\bibnamefont {Kaldun}},
  \bibinfo {author} {\bibfnamefont {P.}~\bibnamefont {Raith}}, \bibinfo
  {author} {\bibfnamefont {K.}~\bibnamefont {Meyer}}, \bibinfo {author}
  {\bibfnamefont {M.}~\bibnamefont {Laux}}, \bibinfo {author} {\bibfnamefont
  {J.}~\bibnamefont {Evers}}, \bibinfo {author} {\bibfnamefont {C.~H.}\
  \bibnamefont {Keitel}}, \bibinfo {author} {\bibfnamefont {C.~H.}\
  \bibnamefont {Greene}},\ and\ \bibinfo {author} {\bibfnamefont
  {T.}~\bibnamefont {Pfeifer}},\ }\bibfield  {title} {\bibinfo {title} {Lorentz
  meets fano in spectral line shapes: A universal phase and its laser
  control},\ }\href {https://doi.org/10.1126/science.1234407} {\bibfield
  {journal} {\bibinfo  {journal} {Science}\ }\textbf {\bibinfo {volume}
  {340}},\ \bibinfo {pages} {716} (\bibinfo {year} {2013})}\BibitemShut
  {NoStop}%
\bibitem [{\citenamefont {Li}\ \emph {et~al.}(2015)\citenamefont {Li},
  \citenamefont {Bernhardt}, \citenamefont {Beck}, \citenamefont {Warrick},
  \citenamefont {Pfeiffer}, \citenamefont {Bell}, \citenamefont {Haxton},
  \citenamefont {McCurdy}, \citenamefont {Neumark},\ and\ \citenamefont
  {Leone}}]{Li2015InvestigationXenon}%
  \BibitemOpen
  \bibfield  {author} {\bibinfo {author} {\bibfnamefont {X.}~\bibnamefont
  {Li}}, \bibinfo {author} {\bibfnamefont {B.}~\bibnamefont {Bernhardt}},
  \bibinfo {author} {\bibfnamefont {A.~R.}\ \bibnamefont {Beck}}, \bibinfo
  {author} {\bibfnamefont {E.~R.}\ \bibnamefont {Warrick}}, \bibinfo {author}
  {\bibfnamefont {A.~N.}\ \bibnamefont {Pfeiffer}}, \bibinfo {author}
  {\bibfnamefont {M.~J.}\ \bibnamefont {Bell}}, \bibinfo {author}
  {\bibfnamefont {D.~J.}\ \bibnamefont {Haxton}}, \bibinfo {author}
  {\bibfnamefont {C.~W.}\ \bibnamefont {McCurdy}}, \bibinfo {author}
  {\bibfnamefont {D.~M.}\ \bibnamefont {Neumark}},\ and\ \bibinfo {author}
  {\bibfnamefont {S.~R.}\ \bibnamefont {Leone}},\ }\bibfield  {title} {\bibinfo
  {title} {Investigation of coupling mechanisms in attosecond transient
  absorption of autoionizing states: comparison of theory and experiment in
  xenon},\ }\href {https://doi.org/10.1088/0953-4075/48/12/125601} {\bibfield
  {journal} {\bibinfo  {journal} {Journal of Physics B: Atomic, Molecular and
  Optical Physics}\ }\textbf {\bibinfo {volume} {48}},\ \bibinfo {pages}
  {125601} (\bibinfo {year} {2015})}\BibitemShut {NoStop}%
\bibitem [{\citenamefont {Kaldun}\ \emph {et~al.}(2016)\citenamefont {Kaldun},
  \citenamefont {Blättermann}, \citenamefont {Stooß}, \citenamefont {Donsa},
  \citenamefont {Wei}, \citenamefont {Pazourek}, \citenamefont {Nagele},
  \citenamefont {Ott}, \citenamefont {Lin}, \citenamefont {Burgdörfer},\ and\
  \citenamefont {Pfeifer}}]{Kaldun2016ObservingDomain}%
  \BibitemOpen
  \bibfield  {author} {\bibinfo {author} {\bibfnamefont {A.}~\bibnamefont
  {Kaldun}}, \bibinfo {author} {\bibfnamefont {A.}~\bibnamefont
  {Blättermann}}, \bibinfo {author} {\bibfnamefont {V.}~\bibnamefont
  {Stooß}}, \bibinfo {author} {\bibfnamefont {S.}~\bibnamefont {Donsa}},
  \bibinfo {author} {\bibfnamefont {H.}~\bibnamefont {Wei}}, \bibinfo {author}
  {\bibfnamefont {R.}~\bibnamefont {Pazourek}}, \bibinfo {author}
  {\bibfnamefont {S.}~\bibnamefont {Nagele}}, \bibinfo {author} {\bibfnamefont
  {C.}~\bibnamefont {Ott}}, \bibinfo {author} {\bibfnamefont {C.~D.}\
  \bibnamefont {Lin}}, \bibinfo {author} {\bibfnamefont {J.}~\bibnamefont
  {Burgdörfer}},\ and\ \bibinfo {author} {\bibfnamefont {T.}~\bibnamefont
  {Pfeifer}},\ }\bibfield  {title} {\bibinfo {title} {Observing the ultrafast
  buildup of a fano resonance in the time domain},\ }\href
  {https://doi.org/10.1126/science.aah6972} {\bibfield  {journal} {\bibinfo
  {journal} {Science}\ }\textbf {\bibinfo {volume} {354}},\ \bibinfo {pages}
  {738} (\bibinfo {year} {2016})}\BibitemShut {NoStop}%
\bibitem [{\citenamefont {Liao}\ \emph {et~al.}(2017)\citenamefont {Liao},
  \citenamefont {Li}, \citenamefont {Haxton}, \citenamefont {Rescigno},
  \citenamefont {Lucchese}, \citenamefont {McCurdy},\ and\ \citenamefont
  {Sandhu}}]{Liao2017Oxygen}%
  \BibitemOpen
  \bibfield  {author} {\bibinfo {author} {\bibfnamefont {C.-T.}\ \bibnamefont
  {Liao}}, \bibinfo {author} {\bibfnamefont {X.}~\bibnamefont {Li}}, \bibinfo
  {author} {\bibfnamefont {D.~J.}\ \bibnamefont {Haxton}}, \bibinfo {author}
  {\bibfnamefont {T.~N.}\ \bibnamefont {Rescigno}}, \bibinfo {author}
  {\bibfnamefont {R.~R.}\ \bibnamefont {Lucchese}}, \bibinfo {author}
  {\bibfnamefont {C.~W.}\ \bibnamefont {McCurdy}},\ and\ \bibinfo {author}
  {\bibfnamefont {A.}~\bibnamefont {Sandhu}},\ }\bibfield  {title} {\bibinfo
  {title} {Probing autoionizing states of molecular oxygen with xuv transient
  absorption: Electronic-symmetry-dependent line shapes and laser-induced
  modifications},\ }\href {https://doi.org/10.1103/PhysRevA.95.043427}
  {\bibfield  {journal} {\bibinfo  {journal} {Phys. Rev. A}\ }\textbf {\bibinfo
  {volume} {95}},\ \bibinfo {pages} {043427} (\bibinfo {year}
  {2017})}\BibitemShut {NoStop}%
\bibitem [{\citenamefont {H{\"{u}}tten}\ \emph {et~al.}(2018)\citenamefont
  {H{\"{u}}tten}, \citenamefont {Mittermair}, \citenamefont {Stock},
  \citenamefont {Beerwerth}, \citenamefont {Shirvanyan}, \citenamefont
  {Riemensberger}, \citenamefont {Duensing}, \citenamefont {Heider},
  \citenamefont {Wagner}, \citenamefont {Guggenmos}, \citenamefont {Fritzsche},
  \citenamefont {Kabachnik}, \citenamefont {Kienberger},\ and\ \citenamefont
  {Bernhardt}}]{Hutten2018UltrafastKrypton}%
  \BibitemOpen
  \bibfield  {author} {\bibinfo {author} {\bibfnamefont {K.}~\bibnamefont
  {H{\"{u}}tten}}, \bibinfo {author} {\bibfnamefont {M.}~\bibnamefont
  {Mittermair}}, \bibinfo {author} {\bibfnamefont {S.~O.}\ \bibnamefont
  {Stock}}, \bibinfo {author} {\bibfnamefont {R.}~\bibnamefont {Beerwerth}},
  \bibinfo {author} {\bibfnamefont {V.}~\bibnamefont {Shirvanyan}}, \bibinfo
  {author} {\bibfnamefont {J.}~\bibnamefont {Riemensberger}}, \bibinfo {author}
  {\bibfnamefont {A.}~\bibnamefont {Duensing}}, \bibinfo {author}
  {\bibfnamefont {R.}~\bibnamefont {Heider}}, \bibinfo {author} {\bibfnamefont
  {M.~S.}\ \bibnamefont {Wagner}}, \bibinfo {author} {\bibfnamefont
  {A.}~\bibnamefont {Guggenmos}}, \bibinfo {author} {\bibfnamefont
  {S.}~\bibnamefont {Fritzsche}}, \bibinfo {author} {\bibfnamefont {N.~M.}\
  \bibnamefont {Kabachnik}}, \bibinfo {author} {\bibfnamefont {R.}~\bibnamefont
  {Kienberger}},\ and\ \bibinfo {author} {\bibfnamefont {B.}~\bibnamefont
  {Bernhardt}},\ }\bibfield  {title} {\bibinfo {title} {{Ultrafast quantum
  control of ionization dynamics in krypton}},\ }\href
  {https://doi.org/10.1038/s41467-018-03122-1} {\bibfield  {journal} {\bibinfo
  {journal} {Nature Communications}\ }\textbf {\bibinfo {volume} {9}},\
  \bibinfo {pages} {1} (\bibinfo {year} {2018})}\BibitemShut {NoStop}%
\bibitem [{\citenamefont {Wu}\ \emph {et~al.}(2016)\citenamefont {Wu},
  \citenamefont {Chen}, \citenamefont {Camp}, \citenamefont {Schafer},\ and\
  \citenamefont {Gaarde}}]{Wu2016TheoryAbsorption}%
  \BibitemOpen
  \bibfield  {author} {\bibinfo {author} {\bibfnamefont {M.}~\bibnamefont
  {Wu}}, \bibinfo {author} {\bibfnamefont {S.}~\bibnamefont {Chen}}, \bibinfo
  {author} {\bibfnamefont {S.}~\bibnamefont {Camp}}, \bibinfo {author}
  {\bibfnamefont {K.~J.}\ \bibnamefont {Schafer}},\ and\ \bibinfo {author}
  {\bibfnamefont {M.~B.}\ \bibnamefont {Gaarde}},\ }\bibfield  {title}
  {\bibinfo {title} {Theory of strong-field attosecond transient absorption},\
  }\href {https://doi.org/10.1088/0953-4075/49/6/062003} {\bibfield  {journal}
  {\bibinfo  {journal} {Journal of Physics B: Atomic, Molecular and Optical
  Physics}\ }\textbf {\bibinfo {volume} {49}},\ \bibinfo {pages} {062003}
  (\bibinfo {year} {2016})}\BibitemShut {NoStop}%
\bibitem [{\citenamefont {Lambropoulos}\ and\ \citenamefont
  {Zoller}(1981)}]{Lambropoulos1981AutoionizingFields}%
  \BibitemOpen
  \bibfield  {author} {\bibinfo {author} {\bibfnamefont {P.}~\bibnamefont
  {Lambropoulos}}\ and\ \bibinfo {author} {\bibfnamefont {P.}~\bibnamefont
  {Zoller}},\ }\bibfield  {title} {\bibinfo {title} {Autoionizing states in
  strong laser fields},\ }\href {https://doi.org/10.1103/PhysRevA.24.379}
  {\bibfield  {journal} {\bibinfo  {journal} {Phys. Rev. A}\ }\textbf {\bibinfo
  {volume} {24}},\ \bibinfo {pages} {379} (\bibinfo {year} {1981})}\BibitemShut
  {NoStop}%
\bibitem [{\citenamefont {Kim}\ and\ \citenamefont
  {Lambropoulos}(1982)}]{Kim1982Laser-IntensityAutoionization}%
  \BibitemOpen
  \bibfield  {author} {\bibinfo {author} {\bibfnamefont {Y.~S.}\ \bibnamefont
  {Kim}}\ and\ \bibinfo {author} {\bibfnamefont {P.}~\bibnamefont
  {Lambropoulos}},\ }\bibfield  {title} {\bibinfo {title} {{Laser-Intensity
  Effect on the Configuration Interaction in Multiphoton Ionization and
  Autoionization}},\ }\href {https://doi.org/10.1103/PhysRevLett.18.301}
  {\bibfield  {journal} {\bibinfo  {journal} {Physical Review Letters}\
  }\textbf {\bibinfo {volume} {49}},\ \bibinfo {pages} {1698} (\bibinfo {year}
  {1982})}\BibitemShut {NoStop}%
\bibitem [{\citenamefont {Chen}\ \emph {et~al.}(2012)\citenamefont {Chen},
  \citenamefont {Bell}, \citenamefont {Beck}, \citenamefont {Mashiko},
  \citenamefont {Wu}, \citenamefont {Pfeiffer}, \citenamefont {Gaarde},
  \citenamefont {Neumark}, \citenamefont {Leone},\ and\ \citenamefont
  {Schafer}}]{Chen2012Helium}%
  \BibitemOpen
  \bibfield  {author} {\bibinfo {author} {\bibfnamefont {S.}~\bibnamefont
  {Chen}}, \bibinfo {author} {\bibfnamefont {M.~J.}\ \bibnamefont {Bell}},
  \bibinfo {author} {\bibfnamefont {A.~R.}\ \bibnamefont {Beck}}, \bibinfo
  {author} {\bibfnamefont {H.}~\bibnamefont {Mashiko}}, \bibinfo {author}
  {\bibfnamefont {M.}~\bibnamefont {Wu}}, \bibinfo {author} {\bibfnamefont
  {A.~N.}\ \bibnamefont {Pfeiffer}}, \bibinfo {author} {\bibfnamefont {M.~B.}\
  \bibnamefont {Gaarde}}, \bibinfo {author} {\bibfnamefont {D.~M.}\
  \bibnamefont {Neumark}}, \bibinfo {author} {\bibfnamefont {S.~R.}\
  \bibnamefont {Leone}},\ and\ \bibinfo {author} {\bibfnamefont {K.~J.}\
  \bibnamefont {Schafer}},\ }\bibfield  {title} {\bibinfo {title}
  {Light-induced states in attosecond transient absorption spectra of
  laser-dressed helium},\ }\href {https://doi.org/10.1103/PhysRevA.86.063408}
  {\bibfield  {journal} {\bibinfo  {journal} {Phys. Rev. A}\ }\textbf {\bibinfo
  {volume} {86}},\ \bibinfo {pages} {063408} (\bibinfo {year}
  {2012})}\BibitemShut {NoStop}%
\bibitem [{\citenamefont {Harkema}\ \emph {et~al.}(2021)\citenamefont
  {Harkema}, \citenamefont {Cariker}, \citenamefont {Lindroth}, \citenamefont
  {Argenti},\ and\ \citenamefont {Sandhu}}]{Harkema2021AutoionizingIonization}%
  \BibitemOpen
  \bibfield  {author} {\bibinfo {author} {\bibfnamefont {N.}~\bibnamefont
  {Harkema}}, \bibinfo {author} {\bibfnamefont {C.}~\bibnamefont {Cariker}},
  \bibinfo {author} {\bibfnamefont {E.}~\bibnamefont {Lindroth}}, \bibinfo
  {author} {\bibfnamefont {L.}~\bibnamefont {Argenti}},\ and\ \bibinfo {author}
  {\bibfnamefont {A.}~\bibnamefont {Sandhu}},\ }\bibfield  {title} {\bibinfo
  {title} {{Autoionizing Polaritons in Attosecond Atomic Ionization}},\ }\href
  {https://doi.org/10.1103/PhysRevLett.127.023202} {\bibfield  {journal}
  {\bibinfo  {journal} {Physical Review Letters}\ }\textbf {\bibinfo {volume}
  {127}},\ \bibinfo {pages} {23202} (\bibinfo {year} {2021})}\BibitemShut
  {NoStop}%
\bibitem [{\citenamefont {Loh}\ \emph {et~al.}(2008)\citenamefont {Loh},
  \citenamefont {Greene},\ and\ \citenamefont {Leone}}]{ZHLoh2008ATS}%
  \BibitemOpen
  \bibfield  {author} {\bibinfo {author} {\bibfnamefont {Z.-H.}\ \bibnamefont
  {Loh}}, \bibinfo {author} {\bibfnamefont {C.~H.}\ \bibnamefont {Greene}},\
  and\ \bibinfo {author} {\bibfnamefont {S.~R.}\ \bibnamefont {Leone}},\
  }\bibfield  {title} {\bibinfo {title} {Femtosecond induced transparency and
  absorption in the extreme ultraviolet by coherent coupling of the he 2s2p
  (1po) and 2p2 (1se) double excitation states with 800nm light},\ }\href
  {https://doi.org/https://doi.org/10.1016/j.chemphys.2007.11.005} {\bibfield
  {journal} {\bibinfo  {journal} {Chemical Physics}\ }\textbf {\bibinfo
  {volume} {350}},\ \bibinfo {pages} {7} (\bibinfo {year} {2008})},\ \bibinfo
  {note} {femtochemistry and Femtobiology}\BibitemShut {NoStop}%
\bibitem [{\citenamefont {Pfeiffer}\ and\ \citenamefont
  {Leone}(2012)}]{Pfeiffer2012ATS}%
  \BibitemOpen
  \bibfield  {author} {\bibinfo {author} {\bibfnamefont {A.~N.}\ \bibnamefont
  {Pfeiffer}}\ and\ \bibinfo {author} {\bibfnamefont {S.~R.}\ \bibnamefont
  {Leone}},\ }\bibfield  {title} {\bibinfo {title} {Transmission of an isolated
  attosecond pulse in a strong-field dressed atom},\ }\href
  {https://doi.org/10.1103/PhysRevA.85.053422} {\bibfield  {journal} {\bibinfo
  {journal} {Phys. Rev. A}\ }\textbf {\bibinfo {volume} {85}},\ \bibinfo
  {pages} {053422} (\bibinfo {year} {2012})}\BibitemShut {NoStop}%
\bibitem [{\citenamefont {Chini}\ \emph {et~al.}(2013)\citenamefont {Chini},
  \citenamefont {Wang}, \citenamefont {Cheng}, \citenamefont {Wu},
  \citenamefont {Zhao}, \citenamefont {Telnov}, \citenamefont {Chu},\ and\
  \citenamefont {Chang}}]{Chini2013ATS}%
  \BibitemOpen
  \bibfield  {author} {\bibinfo {author} {\bibfnamefont {M.}~\bibnamefont
  {Chini}}, \bibinfo {author} {\bibfnamefont {X.}~\bibnamefont {Wang}},
  \bibinfo {author} {\bibfnamefont {Y.}~\bibnamefont {Cheng}}, \bibinfo
  {author} {\bibfnamefont {Y.}~\bibnamefont {Wu}}, \bibinfo {author}
  {\bibfnamefont {D.}~\bibnamefont {Zhao}}, \bibinfo {author} {\bibfnamefont
  {D.~A.}\ \bibnamefont {Telnov}}, \bibinfo {author} {\bibfnamefont {S.-I.}\
  \bibnamefont {Chu}},\ and\ \bibinfo {author} {\bibfnamefont {Z.}~\bibnamefont
  {Chang}},\ }\bibfield  {title} {\bibinfo {title} {Sub-cycle oscillations in
  virtual states brought to light},\ }\href@noop {} {\bibfield  {journal}
  {\bibinfo  {journal} {Scientific reports}\ }\textbf {\bibinfo {volume} {3}},\
  \bibinfo {pages} {1} (\bibinfo {year} {2013})}\BibitemShut {NoStop}%
\bibitem [{\citenamefont {Wu}\ \emph {et~al.}(2013)\citenamefont {Wu},
  \citenamefont {Chen}, \citenamefont {Gaarde},\ and\ \citenamefont
  {Schafer}}]{Wu2013ATS}%
  \BibitemOpen
  \bibfield  {author} {\bibinfo {author} {\bibfnamefont {M.}~\bibnamefont
  {Wu}}, \bibinfo {author} {\bibfnamefont {S.}~\bibnamefont {Chen}}, \bibinfo
  {author} {\bibfnamefont {M.~B.}\ \bibnamefont {Gaarde}},\ and\ \bibinfo
  {author} {\bibfnamefont {K.~J.}\ \bibnamefont {Schafer}},\ }\bibfield
  {title} {\bibinfo {title} {Time-domain perspective on autler-townes splitting
  in attosecond transient absorption of laser-dressed helium atoms},\ }\href
  {https://doi.org/10.1103/PhysRevA.88.043416} {\bibfield  {journal} {\bibinfo
  {journal} {Phys. Rev. A}\ }\textbf {\bibinfo {volume} {88}},\ \bibinfo
  {pages} {043416} (\bibinfo {year} {2013})}\BibitemShut {NoStop}%
\bibitem [{\citenamefont {Chini}\ \emph {et~al.}(2014)\citenamefont {Chini},
  \citenamefont {Wang}, \citenamefont {Cheng},\ and\ \citenamefont
  {Chang}}]{Chini2014ATS}%
  \BibitemOpen
  \bibfield  {author} {\bibinfo {author} {\bibfnamefont {M.}~\bibnamefont
  {Chini}}, \bibinfo {author} {\bibfnamefont {X.}~\bibnamefont {Wang}},
  \bibinfo {author} {\bibfnamefont {Y.}~\bibnamefont {Cheng}},\ and\ \bibinfo
  {author} {\bibfnamefont {Z.}~\bibnamefont {Chang}},\ }\bibfield  {title}
  {\bibinfo {title} {Resonance effects and quantum beats in attosecond
  transient absorption of helium},\ }\href
  {https://doi.org/10.1088/0953-4075/47/12/124009} {\bibfield  {journal}
  {\bibinfo  {journal} {Journal of Physics B: Atomic, Molecular and Optical
  Physics}\ }\textbf {\bibinfo {volume} {47}},\ \bibinfo {pages} {124009}
  (\bibinfo {year} {2014})}\BibitemShut {NoStop}%
\bibitem [{\citenamefont {Argenti}\ \emph {et~al.}(2015)\citenamefont
  {Argenti}, \citenamefont {Jim\'enez-Gal\'an}, \citenamefont {Marante},
  \citenamefont {Ott}, \citenamefont {Pfeifer},\ and\ \citenamefont
  {Mart\'{\i}n}}]{Argenti2015ATS}%
  \BibitemOpen
  \bibfield  {author} {\bibinfo {author} {\bibfnamefont {L.}~\bibnamefont
  {Argenti}}, \bibinfo {author} {\bibfnamefont {A.}~\bibnamefont
  {Jim\'enez-Gal\'an}}, \bibinfo {author} {\bibfnamefont {C.}~\bibnamefont
  {Marante}}, \bibinfo {author} {\bibfnamefont {C.}~\bibnamefont {Ott}},
  \bibinfo {author} {\bibfnamefont {T.}~\bibnamefont {Pfeifer}},\ and\ \bibinfo
  {author} {\bibfnamefont {F.}~\bibnamefont {Mart\'{\i}n}},\ }\bibfield
  {title} {\bibinfo {title} {Dressing effects in the attosecond transient
  absorption spectra of doubly excited states in helium},\ }\href
  {https://doi.org/10.1103/PhysRevA.91.061403} {\bibfield  {journal} {\bibinfo
  {journal} {Phys. Rev. A}\ }\textbf {\bibinfo {volume} {91}},\ \bibinfo
  {pages} {061403} (\bibinfo {year} {2015})}\BibitemShut {NoStop}%
\bibitem [{\citenamefont {Kobayashi}\ \emph {et~al.}(2017)\citenamefont
  {Kobayashi}, \citenamefont {Timmers}, \citenamefont {Sabbar}, \citenamefont
  {Leone},\ and\ \citenamefont {Neumark}}]{Kobayashi2017ATS}%
  \BibitemOpen
  \bibfield  {author} {\bibinfo {author} {\bibfnamefont {Y.}~\bibnamefont
  {Kobayashi}}, \bibinfo {author} {\bibfnamefont {H.}~\bibnamefont {Timmers}},
  \bibinfo {author} {\bibfnamefont {M.}~\bibnamefont {Sabbar}}, \bibinfo
  {author} {\bibfnamefont {S.~R.}\ \bibnamefont {Leone}},\ and\ \bibinfo
  {author} {\bibfnamefont {D.~M.}\ \bibnamefont {Neumark}},\ }\bibfield
  {title} {\bibinfo {title} {Attosecond transient-absorption dynamics of xenon
  core-excited states in a strong driving field},\ }\href
  {https://doi.org/10.1103/PhysRevA.95.031401} {\bibfield  {journal} {\bibinfo
  {journal} {Phys. Rev. A}\ }\textbf {\bibinfo {volume} {95}},\ \bibinfo
  {pages} {031401} (\bibinfo {year} {2017})}\BibitemShut {NoStop}%
\bibitem [{\citenamefont {Harkema}\ \emph {et~al.}(2018)\citenamefont
  {Harkema}, \citenamefont {B{\ae}kh{\o}j}, \citenamefont {Liao}, \citenamefont
  {Gaarde}, \citenamefont {Schafer},\ and\ \citenamefont
  {Sandhu}}]{Harkema2018Noncommensurate}%
  \BibitemOpen
  \bibfield  {author} {\bibinfo {author} {\bibfnamefont {N.}~\bibnamefont
  {Harkema}}, \bibinfo {author} {\bibfnamefont {J.~E.}\ \bibnamefont
  {B{\ae}kh{\o}j}}, \bibinfo {author} {\bibfnamefont {C.-T.}\ \bibnamefont
  {Liao}}, \bibinfo {author} {\bibfnamefont {M.~B.}\ \bibnamefont {Gaarde}},
  \bibinfo {author} {\bibfnamefont {K.~J.}\ \bibnamefont {Schafer}},\ and\
  \bibinfo {author} {\bibfnamefont {A.}~\bibnamefont {Sandhu}},\ }\bibfield
  {title} {\bibinfo {title} {Controlling attosecond transient absorption with
  tunable, non-commensurate light fields},\ }\href
  {https://doi.org/10.1364/OL.43.003357} {\bibfield  {journal} {\bibinfo
  {journal} {Opt. Lett.}\ }\textbf {\bibinfo {volume} {43}},\ \bibinfo {pages}
  {3357} (\bibinfo {year} {2018})}\BibitemShut {NoStop}%
\bibitem [{\citenamefont {Corkum}(1993)}]{Corkum1993PlasmaIonization}%
  \BibitemOpen
  \bibfield  {author} {\bibinfo {author} {\bibfnamefont {P.~B.}\ \bibnamefont
  {Corkum}},\ }\bibfield  {title} {\bibinfo {title} {{Plasma perspective on
  strong field multiphoton ionization}},\ }\href
  {https://doi.org/10.1103/PhysRevLett.71.1994} {\bibfield  {journal} {\bibinfo
   {journal} {Physical Review Letters}\ }\textbf {\bibinfo {volume} {71}},\
  \bibinfo {pages} {1994} (\bibinfo {year} {1993})}\BibitemShut {NoStop}%
\bibitem [{\citenamefont {Antoine}\ \emph {et~al.}(1996)\citenamefont
  {Antoine}, \citenamefont {L’huillier},\ and\ \citenamefont
  {Lewenstein}}]{Antoine1996AttosecondHarmonics}%
  \BibitemOpen
  \bibfield  {author} {\bibinfo {author} {\bibfnamefont {P.}~\bibnamefont
  {Antoine}}, \bibinfo {author} {\bibfnamefont {A.}~\bibnamefont
  {L’huillier}},\ and\ \bibinfo {author} {\bibfnamefont {M.}~\bibnamefont
  {Lewenstein}},\ }\bibfield  {title} {\bibinfo {title} {{Attosecond pulse
  trains using high–order harmonics}},\ }\href
  {https://doi.org/10.1103/PhysRevLett.77.1234} {\bibfield  {journal} {\bibinfo
   {journal} {Physical Review Letters}\ }\textbf {\bibinfo {volume} {77}},\
  \bibinfo {pages} {1234} (\bibinfo {year} {1996})}\BibitemShut {NoStop}%
\bibitem [{\citenamefont {Paul}\ \emph {et~al.}(2001)\citenamefont {Paul},
  \citenamefont {Toma}, \citenamefont {Breger}, \citenamefont {Mullot},
  \citenamefont {Aug{\'{e}}}, \citenamefont {Balcou}, \citenamefont {Muller},\
  and\ \citenamefont {Agostini}}]{Paul2001ObservationGeneration}%
  \BibitemOpen
  \bibfield  {author} {\bibinfo {author} {\bibfnamefont {P.~M.}\ \bibnamefont
  {Paul}}, \bibinfo {author} {\bibfnamefont {E.~S.}\ \bibnamefont {Toma}},
  \bibinfo {author} {\bibfnamefont {P.}~\bibnamefont {Breger}}, \bibinfo
  {author} {\bibfnamefont {G.}~\bibnamefont {Mullot}}, \bibinfo {author}
  {\bibfnamefont {F.}~\bibnamefont {Aug{\'{e}}}}, \bibinfo {author}
  {\bibfnamefont {P.}~\bibnamefont {Balcou}}, \bibinfo {author} {\bibfnamefont
  {H.~G.}\ \bibnamefont {Muller}},\ and\ \bibinfo {author} {\bibfnamefont
  {P.}~\bibnamefont {Agostini}},\ }\bibfield  {title} {\bibinfo {title}
  {{Observation of a train of attosecond pulses from high harmonic
  generation}},\ }\href {https://doi.org/10.1126/science.1059413} {\bibfield
  {journal} {\bibinfo  {journal} {Science}\ }\textbf {\bibinfo {volume}
  {292}},\ \bibinfo {pages} {1689} (\bibinfo {year} {2001})}\BibitemShut
  {NoStop}%
\bibitem [{\citenamefont {Carette}\ \emph {et~al.}(2013)\citenamefont
  {Carette}, \citenamefont {Dahlstr{\"{o}}m}, \citenamefont {Argenti},\ and\
  \citenamefont {Lindroth}}]{Carette2013}%
  \BibitemOpen
  \bibfield  {author} {\bibinfo {author} {\bibfnamefont {T.}~\bibnamefont
  {Carette}}, \bibinfo {author} {\bibfnamefont {J.~M.}\ \bibnamefont
  {Dahlstr{\"{o}}m}}, \bibinfo {author} {\bibfnamefont {L.}~\bibnamefont
  {Argenti}},\ and\ \bibinfo {author} {\bibfnamefont {E.}~\bibnamefont
  {Lindroth}},\ }\bibfield  {title} {\bibinfo {title} {{Multiconfigurational
  Hartree-Fock close-coupling ansatz: Application to the argon photoionization
  cross section and delays}},\ }\href
  {https://doi.org/10.1103/PhysRevA.87.023420} {\bibfield  {journal} {\bibinfo
  {journal} {Phys. Rev. A}\ }\textbf {\bibinfo {volume} {87}},\ \bibinfo
  {pages} {023420} (\bibinfo {year} {2013})}\BibitemShut {NoStop}%
\bibitem [{\citenamefont {Marante}\ \emph {et~al.}(2017)\citenamefont
  {Marante}, \citenamefont {Klinker}, \citenamefont {Kjellsson}, \citenamefont
  {Lindroth}, \citenamefont {Gonz\'alez-V\'azquez}, \citenamefont {Argenti},\
  and\ \citenamefont {Mart\'{\i}n}}]{Marante2017}%
  \BibitemOpen
  \bibfield  {author} {\bibinfo {author} {\bibfnamefont {C.}~\bibnamefont
  {Marante}}, \bibinfo {author} {\bibfnamefont {M.}~\bibnamefont {Klinker}},
  \bibinfo {author} {\bibfnamefont {T.}~\bibnamefont {Kjellsson}}, \bibinfo
  {author} {\bibfnamefont {E.}~\bibnamefont {Lindroth}}, \bibinfo {author}
  {\bibfnamefont {J.}~\bibnamefont {Gonz\'alez-V\'azquez}}, \bibinfo {author}
  {\bibfnamefont {L.}~\bibnamefont {Argenti}},\ and\ \bibinfo {author}
  {\bibfnamefont {F.}~\bibnamefont {Mart\'{\i}n}},\ }\bibfield  {title}
  {\bibinfo {title} {Photoionization using the xchem approach: Total and
  partial cross sections of ne and resonance parameters above the
  $2{s}^{2}2{p}^{5}$ threshold},\ }\href
  {https://doi.org/10.1103/PhysRevA.96.022507} {\bibfield  {journal} {\bibinfo
  {journal} {Phys. Rev. A}\ }\textbf {\bibinfo {volume} {96}},\ \bibinfo
  {pages} {022507} (\bibinfo {year} {2017})}\BibitemShut {NoStop}%
\bibitem [{\citenamefont {Chew}\ \emph {et~al.}(2018)\citenamefont {Chew},
  \citenamefont {Douguet}, \citenamefont {Cariker}, \citenamefont {Li},
  \citenamefont {Lindroth}, \citenamefont {Ren}, \citenamefont {Yin},
  \citenamefont {Argenti}, \citenamefont {Hill},\ and\ \citenamefont
  {Chang}}]{Chew2018}%
  \BibitemOpen
  \bibfield  {author} {\bibinfo {author} {\bibfnamefont {A.}~\bibnamefont
  {Chew}}, \bibinfo {author} {\bibfnamefont {N.}~\bibnamefont {Douguet}},
  \bibinfo {author} {\bibfnamefont {C.}~\bibnamefont {Cariker}}, \bibinfo
  {author} {\bibfnamefont {J.}~\bibnamefont {Li}}, \bibinfo {author}
  {\bibfnamefont {E.}~\bibnamefont {Lindroth}}, \bibinfo {author}
  {\bibfnamefont {X.}~\bibnamefont {Ren}}, \bibinfo {author} {\bibfnamefont
  {Y.}~\bibnamefont {Yin}}, \bibinfo {author} {\bibfnamefont {L.}~\bibnamefont
  {Argenti}}, \bibinfo {author} {\bibfnamefont {W.~T.}\ \bibnamefont {Hill}},\
  and\ \bibinfo {author} {\bibfnamefont {Z.}~\bibnamefont {Chang}},\ }\bibfield
   {title} {\bibinfo {title} {Attosecond transient absorption spectrum of argon
  at the ${L}_{2,3}$ edge},\ }\href
  {https://doi.org/10.1103/PhysRevA.97.031407} {\bibfield  {journal} {\bibinfo
  {journal} {Phys. Rev. A}\ }\textbf {\bibinfo {volume} {97}},\ \bibinfo
  {pages} {031407(R)} (\bibinfo {year} {2018})}\BibitemShut {NoStop}%
\bibitem [{\citenamefont {Froese~Fischer}\ \emph {et~al.}(2007)\citenamefont
  {Froese~Fischer}, \citenamefont {Tachiev}, \citenamefont {Gaigalas},\ and\
  \citenamefont {Godefroid}}]{Froese2007}%
  \BibitemOpen
  \bibfield  {author} {\bibinfo {author} {\bibfnamefont {C.}~\bibnamefont
  {Froese~Fischer}}, \bibinfo {author} {\bibfnamefont {G.}~\bibnamefont
  {Tachiev}}, \bibinfo {author} {\bibfnamefont {G.}~\bibnamefont {Gaigalas}},\
  and\ \bibinfo {author} {\bibfnamefont {M.~R.}\ \bibnamefont {Godefroid}},\
  }\bibfield  {title} {\bibinfo {title} {An {MCHF} atomic-structure package for
  large-scale calculations},\ }\href
  {https://doi.org/10.1016/j.cpc.2007.01.006} {\bibfield  {journal} {\bibinfo
  {journal} {Comp. Phys. Commun.}\ }\textbf {\bibinfo {volume} {176}},\
  \bibinfo {pages} {559} (\bibinfo {year} {2007})}\BibitemShut {NoStop}%
\bibitem [{\citenamefont {Argenti}\ and\ \citenamefont
  {Moccia}(2016)}]{Argenti2016}%
  \BibitemOpen
  \bibfield  {author} {\bibinfo {author} {\bibfnamefont {L.}~\bibnamefont
  {Argenti}}\ and\ \bibinfo {author} {\bibfnamefont {R.}~\bibnamefont
  {Moccia}},\ }\bibfield  {title} {\bibinfo {title} {Autoionizing states of
  atomic boron},\ }\href {https://doi.org/10.1103/PhysRevA.93.042503}
  {\bibfield  {journal} {\bibinfo  {journal} {Phys. Rev. A}\ }\textbf {\bibinfo
  {volume} {93}},\ \bibinfo {pages} {042503} (\bibinfo {year}
  {2016})}\BibitemShut {NoStop}%
\bibitem [{\citenamefont {Siegert}(1939)}]{Siegert1939}%
  \BibitemOpen
  \bibfield  {author} {\bibinfo {author} {\bibfnamefont {A.~J.~F.}\
  \bibnamefont {Siegert}},\ }\bibfield  {title} {\bibinfo {title} {On the
  derivation of the dispersion formula for nuclear reactions},\ }\href
  {https://doi.org/10.1103/PhysRev.56.750} {\bibfield  {journal} {\bibinfo
  {journal} {Phys. Rev.}\ }\textbf {\bibinfo {volume} {56}},\ \bibinfo {pages}
  {750} (\bibinfo {year} {1939})}\BibitemShut {NoStop}%
\bibitem [{\citenamefont {Tolstikhin}\ \emph {et~al.}(1997)\citenamefont
  {Tolstikhin}, \citenamefont {Ostrovsky},\ and\ \citenamefont
  {Nakamura}}]{Tolstikhin1997}%
  \BibitemOpen
  \bibfield  {author} {\bibinfo {author} {\bibfnamefont {O.~I.}\ \bibnamefont
  {Tolstikhin}}, \bibinfo {author} {\bibfnamefont {V.~N.}\ \bibnamefont
  {Ostrovsky}},\ and\ \bibinfo {author} {\bibfnamefont {H.}~\bibnamefont
  {Nakamura}},\ }\bibfield  {title} {\bibinfo {title} {Siegert pseudo-states as
  a universal tool: Resonances, $\mathit{S}$ matrix, green function},\ }\href
  {https://doi.org/10.1103/PhysRevLett.79.2026} {\bibfield  {journal} {\bibinfo
   {journal} {Phys. Rev. Lett.}\ }\textbf {\bibinfo {volume} {79}},\ \bibinfo
  {pages} {2026} (\bibinfo {year} {1997})}\BibitemShut {NoStop}%
\bibitem [{\citenamefont {Tolstikhin}(2006)}]{Tolstikhin2006}%
  \BibitemOpen
  \bibfield  {author} {\bibinfo {author} {\bibfnamefont {O.~I.}\ \bibnamefont
  {Tolstikhin}},\ }\bibfield  {title} {\bibinfo {title} {Siegert-state
  expansion for nonstationary systems: Coupled equations in the one-channel
  case},\ }\href {https://doi.org/10.1103/PhysRevA.73.062705} {\bibfield
  {journal} {\bibinfo  {journal} {Phys. Rev. A}\ }\textbf {\bibinfo {volume}
  {73}},\ \bibinfo {pages} {062705} (\bibinfo {year} {2006})}\BibitemShut
  {NoStop}%
\bibitem [{\citenamefont {Jaynes}\ and\ \citenamefont
  {Cummings}(1963)}]{JCM1963}%
  \BibitemOpen
  \bibfield  {author} {\bibinfo {author} {\bibfnamefont {E.}~\bibnamefont
  {Jaynes}}\ and\ \bibinfo {author} {\bibfnamefont {F.}~\bibnamefont
  {Cummings}},\ }\bibfield  {title} {\bibinfo {title} {Comparison of quantum
  and semiclassical radiation theories with application to the beam maser},\
  }\href {https://doi.org/10.1109/PROC.1963.1664} {\bibfield  {journal}
  {\bibinfo  {journal} {Proceedings of the IEEE}\ }\textbf {\bibinfo {volume}
  {51}},\ \bibinfo {pages} {89} (\bibinfo {year} {1963})}\BibitemShut {NoStop}%
\bibitem [{\citenamefont {Greentree}\ \emph {et~al.}(2013)\citenamefont
  {Greentree}, \citenamefont {Koch},\ and\ \citenamefont
  {Larson}}]{greentree2013fiftyJCM}%
  \BibitemOpen
  \bibfield  {author} {\bibinfo {author} {\bibfnamefont {A.~D.}\ \bibnamefont
  {Greentree}}, \bibinfo {author} {\bibfnamefont {J.}~\bibnamefont {Koch}},\
  and\ \bibinfo {author} {\bibfnamefont {J.}~\bibnamefont {Larson}},\
  }\bibfield  {title} {\bibinfo {title} {Fifty years of jaynes--cummings
  physics},\ }\href
  {https://iopscience.iop.org/article/10.1088/0953-4075/46/22/220201}
  {\bibfield  {journal} {\bibinfo  {journal} {Journal of Physics B: Atomic,
  Molecular and Optical Physics}\ }\textbf {\bibinfo {volume} {46}},\ \bibinfo
  {pages} {220201} (\bibinfo {year} {2013})}\BibitemShut {NoStop}%
\end{thebibliography}%

\end{document}